%% file: main.tex
\documentclass[preprint,numbers,9pt]{sigplanconf}

\usepackage[utf8]{inputenc}
\usepackage{amsmath,
	    amsfonts,
	    amstext,
	    amssymb,
	    mathrsfs,
	    stmaryrd,
	    latexsym,
	    enumitem,
	    proof,
	    graphicx,
	    color,
	    hyperref,
	    comment}

\input{Commands}

\input{Environments}

\newcommand{\PML}{\textrm{PML}}
\newcommand{\PMLo}{\ensuremath{\PML_\vee}}
\newcommand{\PMLa}{\ensuremath{\PML_\wedge}}
\newcommand{\PMLno}{\ensuremath{\PML_{{\neg}{\vee}}}}
\newcommand{\PMLna}{\ensuremath{\PML_{{\neg}{\wedge}}}}
\newcommand{\trs}[1]{\arrow{#1}{}}
\newcommand{\Distr}{\mathop{\textit{D}}}
\newcommand{\cat}{\textsf}
\newcommand{\Set}{\cat{Set}}
\newcommand{\BRP}{B_{RP}}
\newcommand{\defeq}{\triangleq}
\newcommand{\Coalg}{\mathop{\textit{Coalg}}}
\newcommand{\RPT}{\textit{RPT}}
\newcommand{\RPTf}{\textit{RPT}_f}
\newcommand{\pow}{\mathcal{P}}
\newcommand{\nil}{\textsl{nil}}
\newcommand{\dsb}[1]{{\llbracket #1 \rrbracket}}
\newcommand{\pbis}{\sbis{\rm PB}}

\newcommand{\prune}[2]{#2 |_{#1}}
\newcommand{\logeq}[1]{\cong_\mathcal{#1}}
\newcommand{\diam}[2]{\langle #1 \rangle_{#2}}
\newcommand{\height}{\textit{height}}
\newcommand{\depth}{\textit{depth}}
\newcommand{\supp}{\mathop{\textit{supp}}}
\renewcommand{\succ}{\textit{succ}}
\newcommand{\phior}{\Phi_{\lor}}
\newcommand{\phiand}{\Phi_{\land}}
\newcommand{\hplb}{\mathop{\textit{hplb}}}
\newcommand{\splb}{\mathop{\textit{splb}}}
\newcommand{\init}{\mathop{\textit{init}}}

\setlist[itemize]{leftmargin=*}

\begin{document}

\title{Disjunctive Probabilistic Modal Logic is Enough
       for Bisimilarity on Reactive Probabilistic Systems}

\authorinfo{Marco Bernardo}
           {Dip.\ di Scienze Pure e Applicate, \\ Università di Urbino, Italy}{}
\authorinfo{Marino Miculan}
           {Dip.\ di Scienze Matematiche, Informatiche e Fisiche, \\ Università di Udine, Italy}{}

\maketitle

\begin{abstract}
Larsen and Skou characterized probabilistic bisimilarity over reactive probabilistic systems with a logic
including true, negation, conjunction, and a diamond modality decorated with a probabilistic lower bound.
Later on, Desharnais, Edalat, and Panangaden showed that negation is not necessary to characterize the same
equivalence. In this paper, we prove that the logical characterization holds also when conjunction is
replaced by disjunction, with negation still being not necessary. To this end, we introduce \emph{reactive
probabilistic trees}, a fully abstract model for reactive probabilistic systems that allows us to
demonstrate expressiveness of the disjunctive probabilistic modal logic, as well as of the previously
mentioned logics, by means of a compactness argument.
\end{abstract}

\input{intro}
\input{prelim}
\input{categ}
\input{discrim}
\input{concl}

%
%\bibliographystyle{plain}
%\bibliography{lics2016}

\end{document}

%% file: Commands.tex
\def\ms#1{\null\ifmmode\mathord{\mathcode`-="702D\it #1\mathcode`\-="2200}%
	\else$\mathord{\mathcode`-="702D\it #1\mathcode`\-="2200}$\fi}

\newcommand{\cws}[2]
	{\\ \centerline{$#2$} \\[-#1pt]}

\newlength{\spacelen}

\newcommand{\bibtrick}[1]
	{}

\newcommand{\lmp}
	{\{ \! | \,}
\newcommand{\rmp}
	{\, | \! \}}

\newcommand{\calb}
        {\mathcal{B}}

\newcommand{\call}
        {\mathcal{L}}

\newcommand{\calr}
        {\mathcal{R}}

\newcommand{\natns}
	{\mathbb{N}}

\newcommand{\realns}
	{\mathbb{R}}

\newcommand{\arrow}[2]
        {\, {\auxarrow\limits^{#1}}_{#2} \,}
\newcommand{\auxarrow}
	{\mathop{\longrightarrow}}

\newcommand{\sbis}[1]
	{\sim_{#1}}

\newcommand{\rightmodels}
	{\mathrel{= \! \mid}}

\newcommand{\fullbox}
	{{\mbox{}\nolinebreak\hfill{$\rule{2mm}{2mm}$}}}

%% file: Environments.tex
\newtheorem{new_theorem}
	{Theorem}[section]

\newtheorem{new_definition}
	[new_theorem]{Definition}

\newtheorem{new_remark}
	[new_theorem]{Remark}

\newtheorem{new_example}
	[new_theorem]{Example}

\newtheorem{new_lemma}
	[new_theorem]{Lemma}

\newtheorem{new_proposition}
	[new_theorem]{Proposition}

\newtheorem{new_corollary}
	[new_theorem]{Corollary}

\newenvironment{definition}
	{\begin{new_definition}\rm}
	{\end{new_definition}}

\newenvironment{example}
	{\begin{new_example}\rm}
	{\end{new_example}}

\newenvironment{proposition}
	{\begin{new_proposition}\rm}
	{\end{new_proposition}}

\newenvironment{theorem}
	{\begin{new_theorem}\rm}
	{\end{new_theorem}}

\newenvironment{corollary}
	{\begin{new_corollary}\rm}
	{\end{new_corollary}}

\newenvironment{proof}
	{\medskip\noindent{\sc Proof}$\ $}
	{}

%% file: intro.tex
\section{Introduction}
\label{sec:intro}

Since its introduction by Larsen and Skou~\cite{LS91}, \emph{probabilistic bisimilarity} has been used to
compare probabilistic systems. It corresponds to Milner's \emph{strong bisimilarity} for nondeterministic
systems, and coincides with \emph{ordinary lumpability} for Markov chains. Larsen and Skou~\cite{LS91} first
proved that probabilistic bisimilarity for \emph{reactive probabilistic systems} can be given a
\emph{logical characterization}: two processes are bisimilar if and only if they satisfy the same set of
formulas of a propositional modal logic similar to Hennessy-Milner logic~\cite{HM85}. In addition to the
usual constructs $\top$, $\neg$, and $\wedge$, this logic features a diamond modality $\diam{a}{p} \phi$,
which is satisfied by a state if, after performing action~$a$, the probability of being in a state
satisfying $\phi$ is at least~$p$.

Later on, Desharnais, Edalat, and Panangaden~\cite{DEP02} showed that negation is \emph{not} necessary for
discrimination purposes, by working in a \emph{continuous}-state setting. This result has no counterpart in
the nonprobabilistic setting, where Hennessy-Milner logic without negation characterizes \emph{simulation}
equivalence, which is strictly coarser than bisimilarity~\cite{Gla01} (while the two equivalences are known
to coincide on reactive probabilistic processes~\cite{BK00}).

In this paper, we show that \emph{disjunction} can be used in place of conjunction without having to
reintroduce negation. Thus, the constructs $\top$, $\vee$, and $\diam{a}{p}$ suffice to characterize
probabilistic bisimilarity on reactive probabilistic processes. The intuition is that from a conjunctive
distinguishing formula we can often derive a disjunctive one by suitably increasing some probabilistic lower
bounds. Not even this result has a counterpart in the nonprobabilistic setting, where replacing conjunction
with disjunction in the absence of negation yields trace equivalence (this equivalence does \emph{not}
coincide with bisimilarity on reactive probabilistic processes).

The idea of the proof is the following. First, using a simple categorical construction, we show that each
reactive probabilistic process can be given a semantics in a precise canonical form, which we call
\emph{reactive probabilistic tree}. These trees can be seen as the probabilistic counterpart of Winskel's
synchronization trees used for nondeterministic processes. The semantics is \emph{fully abstract}, i.e., two
states are probabilistically bisimilar if and only if they are mapped to the same reactive probabilistic
tree. Moreover, the semantics is \emph{compact}, in the sense that two (possibly infinite) trees are equal
if and only if all of their finite approximations are equal. Hence, in order to prove that our logic
characterizes probabilistic bisimilarity, it suffices to prove that it allows to discriminate \emph{finite}
reactive probabilistic trees. Indeed, given two different finite trees, we show how to construct (by
induction on the height of one of the trees) a \emph{distinguishing formula} of the disjunctive logic that
tells the two trees apart and has a depth not exceeding the height of the two trees. Our technique applies
also to the logics in~\cite{LS91,DEP02}, for which it allows us to provide simpler proofs of expressiveness,
directly in a \emph{discrete} setting. More generally, this technique can be used in any computational model
that has a compact, fully abstract semantics.

\smallskip

\noindent\textit{Synopsis} In Sect.~\ref{sec:prelim}, we recall the basic definitions about reactive
probabilistic processes, bisimilarity, and logics. In Sect.~\ref{sec:categ}, we characterize probabilistic
bisimilarity in terms of finite reactive probabilistic trees. In Sect.~\ref{sec:discrim}, we prove that the
various probabilistic modal logics considered in the paper can discriminate these finite trees, and hence
characterize probabilistic bisimilarity. Conclusions and directions for future work are in
Sect.~\ref{sec:concl}.

%% file: prelim.tex
\section{Processes, Bisimilarity, and Logics}
\label{sec:prelim}
\subsection{Reactive Probabilistic Processes and Strong Bisimilarity}
\label{sec:pts}

Probabilistic processes can be represented as labeled transitions systems~\cite{Kel76} enriched with
probabilistic information used to determine which action is executed or which state is reached. Following
the terminology of~\cite{GSS95}, we focus on \emph{reactive} probabilistic processes, where every state has
for each action at most one outgoing distribution over states; the choice among these arbitrarily many,
differently labeled distributions is nondeterministic. For a countable (i.e., finite or countably infinite)
set $X$, the set of finitely supported (a.k.a.\ simple) probability distributions over $X$ is:
	\begin{equation}\label{eq:distr}
\Distr(X) \: = \: \{ \Delta \! : \! X \! \to \! \realns_{[0,1]} \mid |\supp(\Delta)| \! < \! \omega,
\textstyle \sum\limits_{x \in X} \!\! \Delta(x) \! = \! 1 \}
	\end{equation}
where the \emph{support} is defined as $\supp(\Delta) \defeq \{ x \in X \mid \Delta(x) > 0 \}$.

	\begin{definition}[RPLTS]\label{def:rplts}
A \emph{reactive probabilistic labeled transition system}, RPLTS for short, is a triple $(S, A, \! \trs{}
\!)$ where:
		\begin{itemize}
\item $S$ is a countable set of \emph{states};
\item $A$ is a countable set of \emph{actions};
\item $\! \trs{} \! \subseteq S \times A \times \Distr(S)$ is a \emph{transition relation} such that,
whenever $(s, a, \Delta_1), (s, a, \Delta_2) \in \! \trs{} \!$, then $\Delta_1 = \Delta_2$.
\fullbox
		\end{itemize}
	\end{definition}

An RPLTS can be seen as a directed graph whose edges are labeled by pairs $(a, p) \in A \times \realns_{]0,
1]}$. For every $s \in S$ and $a \in A$, if there are $a$-labeled edges outgoing from $s$, then these are
finitely many (\emph{image finiteness}), because the considered distributions are finitely supported, and
the numbers on them add up to~$1$. As usual, we denote $(s, a, \Delta) \! \in \! \trs{} \!$ as $s \trs{a}
\Delta$, where the set of reachable states coincides with $\supp(\Delta)$. We also define cumulative
reachability as $\Delta(S') = \sum_{s' \in S'} \Delta(s')$ for all $S' \subseteq S$.

Probabilistic bisimilarity for the class of reactive probabilistic processes was introduced by Larsen and
Skou~\cite{LS91}.

	\begin{definition}[Probabilistic bisimilarity]\label{def:bisim}
Let $(S, A, \! \trs{} \!)$ be an RPLTS. An equivalence relation $\calb$ over $S$ is a \emph{probabilistic
bisimulation} iff, whenever $(s_{1}, s_{2}) \in \calb$, then for all actions $a \in A$:
		\begin{itemize}
\item if $s_{1} \arrow{a}{} \Delta_{1}$, then $s_{2} \arrow{a}{} \Delta_{2}$ and $\Delta_{1}(C) =
\Delta_{2}(C)$ for all equivalence classes $C \in S / \calb$;
\item if $s_{2} \arrow{a}{} \Delta_{2}$, then $s_{1} \arrow{a}{} \Delta_{1}$ and $\Delta_{1}(C) =
\Delta_{2}(C)$ for all equivalence classes $C \in S / \calb$.
		\end{itemize}

We say that $s_{1}, s_{2} \in S$ are \emph{probabilistically bisimilar}, written $s_{1} \pbis s_{2}$, iff
there exists a probabilistic bisimulation including the pair $(s_{1}, s_{2})$.
\fullbox
	\end{definition}

\subsection{Probabilistic Modal Logics}
\label{sec:pml}

In our setting, a probabilistic modal logic is a pair formed by a set $\mathcal{L}$ of \emph{formulas} and
an RPLTS-indexed family of \emph{satisfaction relations} $\models \; \subseteq S \times \mathcal{L}$. The
\emph{logical equivalence} induced by $\mathcal{L}$ over $S$ is defined by letting $s_{1} \logeq{L} s_{2}$,
where $s_{1}, s_{2} \in S$, iff $s_{1} \models \phi \iff s_{2} \models \phi$ for all $\phi \in \mathcal{L}$.
We say that $\mathcal{L}$ \emph{characterizes} a binary relation $\calr$ over~$S$ when $\calr = \;
\logeq{L}$.

We are especially interested in probabilistic modal logics characterizing $\pbis$. The logics considered in
this paper are similar to Hennessy-Milner logic~\cite{HM85}, but the diamond modality is decorated with a
probabilistic lower bound as follows:
\[
\begin{array}{lrcl}
\PMLna: \quad & \phi & \!\!\! ::= \!\!\! & \top \mid \neg\phi \mid \phi \wedge \phi \mid \diam{a}{p} \phi \\
\PMLno: \quad & \phi & \!\!\! ::= \!\!\! & \top \mid \neg\phi \mid \phi \vee \phi \mid \diam{a}{p} \phi \\
\PMLa:  \quad & \phi & \!\!\! ::= \!\!\! & \top \mid \phi \wedge \phi \mid \diam{a}{p} \phi \\
\PMLo:  \quad & \phi & \!\!\! ::= \!\!\! & \top \mid \phi \vee \phi \mid \diam{a}{p} \phi \\
\end{array}
\]
where $p \in \realns_{[0, 1]}$; trailing $\top$'s will be omitted for sake of readability. Their semantics
with respect to an RPLTS state $s$ is as usual:
\[
\begin{array}{rcl}
s \models \top & \!\!\! \iff \!\!\! & \text{true} \\
s \models \neg\phi & \!\!\! \iff \!\!\! & s \not\models \phi \\
s \models \phi_{1} \wedge \phi_{2} & \!\!\! \iff \!\!\! & s \models \phi_{1} \text{ and } s \models \phi_{2}
\\
s \models \phi_{1} \vee \phi_{2} & \!\!\! \iff \!\!\! & s \models \phi_{1} \text{ or } s \models \phi_{2} \\
s \models \diam{a}{p} \phi & \!\!\! \iff \!\!\! & s \trs{a} \Delta \text{ and } \Delta(\{ s' \in S \mid s'
\models \phi \}) \geq p \\
\end{array}
\]

Larsen and Skou~\cite{LS91} proved that \PMLna\ characterizes $\pbis$. This holds true for \PMLno\ as well,
because \PMLno\ is equivalent to \PMLna. Desharnais, Edalat, and Panangaden~\cite{DEP02} then proved in a
\emph{measure-theoretic} setting that \PMLa\ characterizes $\pbis$ too, and hence negation is not necessary.
This was subsequently redemonstrated by Jacobs and Sokolova~\cite{JS10} in the \emph{dual adjunction}
framework, as well as by Deng and Wu~\cite{DW14} with a simpler proof. The main aim of this paper is to show
that \PMLo\ suffices as well.

%% file: categ.tex
\section{Compact Characterization of Probabilistic Bisimilarity}
\label{sec:categ}

In this section, we provide a characterization of probabilistic bisimilarity by means of \emph{finite}
structures in a canonical form. To this end, we introduce \emph{reactive probabilistic trees}, a concrete
representation of probabilistic behaviors.

\subsection{Coalgebras for Probabilistic Systems}

We begin by recalling the coalgebraic setting for probabilistic systems; see, e.g., \cite{VR99}.
The function $\Distr$ defined in~\eqref{eq:distr} extends to a functor $\Distr:\Set\to\Set$ whose action on
morphisms is, for $f:X\to Y$:
\[
\Distr(f): \Distr(X)\to \Distr(Y) \qquad \Distr(f)(\Delta) \: = \: \lambda y.\Delta(f^{-1}(y))
\]
Then, it is easy to see that every RPLTS corresponds to a coalgebra of the following functor:
\[
   \BRP : \Set \to \Set \qquad
   \BRP(X) \: = \: (\Distr(X)+1)^A
\]
Indeed, given $S=(S,A, \! \trs{} \!)$, we define the corresponding coalgebra $(S,\sigma)$ as 
\[
\sigma:S\to \BRP(S) \qquad
\sigma(s) \: \defeq \:
\lambda a.\begin{cases} 
					\Delta & \text{if } s\trs{a}\Delta \\
					* & \text{otherwise}
				\end{cases}
\]

A \emph{homomorphism} $h:(S,\sigma)\to(T,\tau)$ is a function $h:S\to T$ which respects the coalgebraic structures, i.e., $\tau\circ h = (\BRP h) \circ \sigma$.
We denote by $\Coalg(\BRP)$ the category of $\BRP$-coalgebras and their homomorphisms.

Aczel and Mendler~\cite{AM89} introduced a general notion of bisimulation for coalgebras, which in our setting instantiates as follows:

	\begin{definition}\label{def:brp_bisim}
Let $(S_1, \sigma_1)$ and $(S_2, \sigma_2)$ be $\BRP$-coalgebras. A relation $\calr \subseteq S_1 \times
S_2$ is a \emph{$\BRP$-bisimulation} iff there exists a coalgebra structure $\rho : \calr \to \BRP \calr$
such that the projections $\pi_1 : \calr \to S_1$ and $\pi_2 : \calr \to S_2$ are homomorphisms (i.e.,
$\sigma_i \circ \pi_i = \BRP \pi_i \circ \rho$ for $i = 1, 2$). 

We say that $s_1 \in S_1$ and $s_2 \in S_2$ are \emph{$\BRP$-bisimilar}, written $s_1 \sim s_2$, iff there
exists a $\BRP$-bisimulation including $(s_1, s_2)$.
\fullbox
	\end{definition}

The following result shows that probabilistic bisimilarity corresponds to $\BRP$-bisimilarity.

	\begin{proposition}\label{prop:bisims}
The probabilistic bisimilarity over an RPLTS $(S, A, \! \trs{} \!)$ coincides with the $\BRP$-bisimilarity
over the corresponding coalgebra $(S, \sigma)$.

		\begin{proof}
An immediate consequence of~\cite[Lemma~4.4 and Thm.~4.5]{VR99}.
\fullbox
		\end{proof}
	\end{proposition}

The next step is to associate each state of a given RPLTS with its \emph{behavior}, i.e., a structure in
some canonical form which we can reason about. These structures can be seen as the elements of the final
coalgebra of $\BRP$, which exists because we consider only finitely supported distributions, as proved in~\cite[Thm.~4.6]{VR99}:

	\begin{proposition}\label{prop:final}
The functor $\BRP$ admits final coalgebra.

		\begin{proof}
The functor $D$ is bounded because it is restricted to distributions with finite support.  Hence also $\BRP$
is bounded; then the final coalgebra exists by the general result~\cite[Thm.~10.4]{Rut00}.
\fullbox
		\end{proof}
	\end{proposition}

Let $(Z,\zeta)$ be a final $\BRP$-coalgebra  (which is unique up-to isomorphism). This coalgebra can be seen as the RPLTS which subsumes all possible behaviors of any RPLTS. Moreover, elements of $Z$ can be seen as ``canonical'' representatives of behaviors, because different states of $Z$ are never bisimilar:

	\begin{proposition}\label{prop:intfulabs}
For all $z_1,z_2\in Z$: $z_1\sim z_2$ iff $z_1=z_2$.
\fullbox
	\end{proposition}

\subsection{Reactive Probabilistic Trees}

Although Prop.~\ref{prop:final} guarantees the existence of the
final coalgebra, it does not provide us with a concrete
representation of its elements. In this subsection, we introduce
\emph{reactive probabilistic trees}, a representation of the final
$\BRP$-coalgebra which can be seen as the natural extension to the
probabilistic setting of \emph{strongly extensional trees} used to
represent the final $\pow_f$-coalgebra~\cite{Wor05}.

	\begin{definition}[\RPT]\label{def:rpt}
An \emph{($A$-labeled) reactive probabilistic tree} is a pair 
$(X,\succ)$ where $X\in\Set$ and $\succ: X\times A \to \pow_f(X\times \realns_{(0,1]})$
are such that the relation $\leq$ over $X$ defined by:
\[
  \frac{}{x\leq x} \qquad \frac{x\leq y \quad z\in\succ(y,a)}{x\leq z}
\]
is a partial order with a least element, called \emph{root},
and for all $x\in X$ and $a\in A$:
\begin{enumerate}
\item the set $\{y\in X\mid y\leq x\}$ is finite and well-ordered;
\item for all $(x_1,p_1),(x_2,p_2) \in \succ(x,a)$, if $x_1 \! = \! x_2$ then $p_1 \! = \! p_2$;
\item for all $(x_1,p_1),(x_2,p_2) \in \succ(x,a)$, if the subtrees rooted at $x_1$ and $x_2$ are isomorphic then $x _1= x_2$;
\item if $\succ(x,a)\neq \emptyset$ then $\sum_{(y,p)\in \succ(x,a)}p=1$.
\end{enumerate}

We denote by $\RPT$, ranged over by $t,t_1,t_2,\dots$, the set of all reactive probabilistic trees (possibly of infinite height), up-to isomorphism.
\fullbox
	\end{definition}

Reactive probabilistic trees are unordered trees where each node for
each action has either no successor or a finite set of successors labeled
with a positive real number adding up to 1; moreover,
subtrees rooted at these successors are all different. See the forthcoming
Fig.~\ref{fig:pml_or_examples} for some examples. In particular, the
trivial tree is $\nil \defeq (\{\bot\},\lambda x,a.\emptyset)$.

For $t=(X,\succ)$, we denote its root by $\bot_t$, its $a$-successors
by $t(a) \defeq succ(\bot_t,a)$, and the subtree rooted at $x\in X$ by
$t[x] \defeq (\{ y \in X \mid x \leq y\}, \lambda y,a.\succ(y,a))$;
thus, $\bot_{t[x]}=x$.

We define $\height : \RPT \to \natns \cup \{ \omega \}$ in the obvious way:
\[
	\height(t) \: \defeq \: \sup \{ 1 + \height(t') \mid (t', p) \in t(a), a \in A \}
\]
where $\sup\emptyset = 0$; hence, $\height(\nil) = 0$.  In particular,
we denote by $\RPTf \defeq \{t\in\RPT \mid \height(t) < \omega \}$ the set
of reactive probabilistic trees of finite height.

A (possibly infinite) tree can be truncated at any height $n$, yielding a finite tree where the missing subtrees are replaced by $\nil$. In order to obtain a $\RPTf$, we need to collapse isomorphic subtrees resulting from the truncation.
More formally, we define first the \emph{truncation} function $tr_n$ by induction on $n$:
$tr_0(t) \defeq \nil$ and 
\begin{align*}
tr_{n+1}(t) & \defeq  ( \{\bot_t\} \cup \bigcup\{ X' \mid ((X',\succ'), p') \in tr_n(t(a)), a\in A\},\\
&\qquad \succ_Y)
\end{align*}
where $\succ_Y(\bot_t,a) \defeq \{ (\bot_{t'},p') \mid (t',p')\in q(t(a))\}$.

The tree returned by $tr_n$ is always finite, but possibly not extensional.
Hence we have to collapse its isomorphic subtrees adding up their weights by means of the $coll$ function as follows:
\begin{align*}
coll(t) & \defeq  ( \{\bot_t\} \cup \bigcup\{ X' \mid ((X',\succ'), p') \in U_a, a\in A\}, \\
 & \qquad \succ_c) \\
\text{where } W_a & =  \{(coll(t'),p)\mid (t',p) \in \succ(\bot_t,a)\} \\
 U_a & \textstyle = \{(s,\sum_{(s,p)\in W_a}p) \mid s\in \pi_1(W_a)\} \\
 succ_c & \textstyle = (\bigcup_{a\in  A} \{(\bot_t,a) \mapsto \{(\bot_s,p)\mid (s,p)\in U_a\}\})
 \cup \\
 &\quad \textstyle \bigcup_{(s,p)\in U_a} \succ_s 
\end{align*}
Finally we can define the \emph{pruning} of $t$ as $\prune{n}{t} \defeq coll(tr_n(t))$.

We have now to show that $\RPT$ is (the carrier of) the final
$\BRP$-coalgebra (up-to isomorphism). In order to simplify the proof,
we reformulate $\BRP$ in a slightly more ``relational'' format.
We define a functor $D':\Set\to\Set$ by letting for any set $X$:
\begin{align*}
D'X & \: = \: \{ U \in \pow_f(X\times \realns_{(0,1]}) \mid 
\textstyle \text{if } U\neq\emptyset \text{ then } \sum_{(x,p)\in U}p=1 \\
 & \hspace{19mm}  \text{and } \forall (x,p),(x',p')\in U: x=x'\Rightarrow p=p' \}
\end{align*}
and for any $f: X\to Y$, the function $D'f :  D'X \to D'Y$ maps
$U\in D'X$ to $\{\textstyle (f(x),\sum_{(x,p)\in U} p) \mid x\in \pi_1(U)\}$.
Then:

	\begin{proposition}\label{prop:equivs}
\begin{enumerate}
\item $D' \cong D+1$.
\item $D'^A \cong \BRP$.
\item $Coalg(D'^A) \cong Coalg(\BRP)$.
\item The supports of the final $D'^A$-coalgebra and of the final $\BRP$-coalgebra are isomorphic.
\end{enumerate}

\begin{proof}
1: For $X\in\Set$, define $\phi_X:D'X \to DX+1$  as 
$\phi_X(\emptyset) = *$, and for $U\neq\emptyset$,
$\phi_X(U): X\to\realns_{[0,1]}$ maps $x$ to~$p$ if $(x,p)\in U$, to $0$ otherwise.
It is easy to check that the $\phi_X$'s are invertible
and form a natural isomorphism $\phi: D' \stackrel{\sim}{\longrightarrow} D+1$. 

2: Trivial by 1; let  $\psi:D'^A \stackrel{\sim}{\longrightarrow}
\BRP$ be the underlying natural isomorphism.  

3: A $D'^A$-coalgebra $(X,\sigma:X\to D'(X)^A)$ is mapped to
$(X,\psi_X\circ \sigma:X\to \BRP(X))$; the vice versa is similar,
using $\psi_X^{-1}$. It is easy to check that these maps are inverse to
each other.

4: Trivial by 3.
\fullbox
\end{proof}
	\end{proposition}

We can now prove that $\RPT$ is the carrier of the final
$\BRP$-coalgebra (up-to isomorphism).  First, we observe that the set $\RPT$ can be
endowed with a $D'^A$-coalgebra structure $\rho:\RPT \to (D'(\RPT))^A$
defined as follows, for $t=(X,\succ)$:
\[
\rho(t)(a) \: \defeq \: \{ (t[x],p) \mid (x,p)\in\succ(\bot_t,a) \}
\]

	\begin{theorem}\label{prop:rptfinal}
$(\RPT,\rho)$ is a final $\BRP$-coalgebra.

\begin{proof}
  By Prop.~\ref{prop:equivs}, it suffices to prove that
  $(\RPT,\rho)$ is the final $D'^A$-coalgebra. To this end, we
  follow the construction given by Worrell in~\cite[Thm.~11]{Wor05}.
  We define an ordinal-indexed
  \emph{final sequence} of sets $(B_\alpha)_\alpha$ together with
  ``projection functions''
  $(f_\gamma^\beta : B_\beta \to B_\gamma)_{\gamma \leq \beta}$:
\begin{alignat*}{3}
B_0 & \: = \: \{\nil\} \cong 1          &  f_0^1 & \: = \:\: ! \\
B_{\alpha+1} & \: = \: D'(B_{\alpha})^A & \quad f_{\alpha+1}^{\alpha+2} & \: = \: D'(f_\alpha^{\alpha+1})^A
\\
B_{\lambda} & \: = \: \lim_{\alpha < \lambda} B_\alpha & f_\alpha^\lambda & \: = \: \pi_\alpha
\quad \text{for $\lambda$ a limit ordinal}
\end{alignat*}
the remaining $f_\gamma^\beta$ being given by suitable compositions.  $D$ is
$\omega$-accessible (because we restrict to finitely supported distributions),
thus by~\cite[Thm.~13]{Wor05} and
Prop.~\ref{prop:equivs} the final sequence converges in
at most $\omega+\omega$ steps to the set $B_{\omega+\omega}$ which
is the carrier of the final $D'^A$-coalgebra.

Now, we have to prove that $B_{\omega+\omega}$ is isomorphic to
$\RPT$.  An element of $B_{\omega+\omega}$ is a sequence of finite 
trees $\vec{t}=(t_0,t_1,\dots)$ such that for each $k \in \omega$
there exists $N_k\in\natns$ such that nodes at depth $k$ of any tree
$t_i$ have at most $N_k$ successors for each label $a\in A$.  These
sequences can be seen as compatible partial views of a single (possibly infinite) tree.
Thus, given a sequence $\vec{t}$ the corresponding tree $u\in\RPT$ is obtained by
\emph{amalgamating} $\vec{t}$: $u$ at depth $k$ is defined by the
level $k$ of a suitable tree $t_i$, where
$i$ is such that for all $j\geq i$, $t_j$ is equal to $t_i$ up to
depth $k$.  On the other hand, given $u\in\RPT$ we can define the
corresponding sequence $\vec{t}\in B_{\omega+\omega}$ as
$t_i = \prune{i}{u}$.

It can be checked that these two maps form an isomorphism between
$B_{\omega+\omega}$ and $\RPT$. Moreover, they respect the coalgebraic
structures, where  
$\tau : B_{\omega+\omega}\to D'(B_{\omega+\omega})^A$ is given by
$\tau(\vec{t})(a) = \{\vec{t'} \in B_{\omega+\omega} \mid \forall i\in\omega:
t'_i\in\succ(t_i,a)\}$.
Therefore, $(B_{\omega+\omega},\tau)$ and $(\RPT,\rho)$ are isomorphic
$D'^A$-coalgebras, hence the thesis.
\fullbox
\end{proof}
	\end{theorem}

\subsection{Full Abstraction and Compactness}

By virtue of Thm.~\ref{prop:rptfinal}, given an RPLTS
$S=(S,A, \! \trs{} \!)$ there exists a unique coalgebra homomorphism
$\dsb{\cdot}:S \to \RPT$, called the \emph{(final) semantics} of
$S$, which associates each state in $S$ with its behavior. 
This semantics is \emph{fully abstract}.

	\begin{theorem}[Full abstraction]\label{thm:fullabs}
Let $(S, A, \! \trs{} \!)$ be an RPLTS. For all $s_{1}, s_{2} \in S$: $s_{1} \pbis s_{2}$ iff
$\dsb{s_{1}} = \dsb{s_{2}}$.

		\begin{proof}
It follows from Props.~\ref{prop:bisims} and~\ref{prop:intfulabs} and Thm.~\ref{prop:rptfinal}.
\fullbox
		\end{proof}
	\end{theorem}

A key property of reactive probabilistic trees is that they are \emph{compact}: two 
different trees can be distinguished by looking at their finite subtrees only. 
Let us formalize this principle:

	\begin{theorem}[Compactness]\label{prop:indprinc}
For all $t_1,t_2\in\RPT$: $t_1=t_2$ iff for all $n\in\natns: \prune{n}{t_1}=\prune{n}{t_2}$.

\begin{proof}
The ``only if'' is trivial. For the ``if'' direction, let us assume that $t_1\neq t_2$; we have to find $n$ such that $\prune{n}{t_1}\neq\prune{n}{t_2}$.
Given a tree $u_0$, a \emph{finite path} in $u_0$ is a sequence $(a_1,p_1,a_2,p_2,\dots,a_n,p_n)$ such that for $i=1,\dots,n: (x_i,p_i)\in u_{i-1}(a_i)$ and $u_i=u[x_i]$.  If $t_1\neq t_2$, then there is a path of length $n$ in, say, $t_1$ which cannot be replayed in $t_2$: in $t_2$ we reach a tree $t'_{n-1}$ such that for all $t$, $(t,p_n)\not\in t'_{n-1}(a_n)$.
Therefore $\prune{n}{t_1}\neq\prune{n}{t_2}$.
\fullbox
\end{proof}
	\end{theorem}

	\begin{corollary}\label{cor:indprinc}
Let $(S, A, \! \trs{} \!)$ be an RPLTS. For all $s_1, s_2 \in S$: $s_1 \pbis s_2$ iff for all $n \in \natns:
\prune{n}{\dsb{s_1}} = \prune{n}{\dsb{s_2}}$.
\fullbox
	\end{corollary}

%% file: discrim.tex
\section{The Discriminating Power of \PMLo}
\label{sec:discrim}

By virtue of the categorical construction leading to Cor.~\ref{cor:indprinc}, in order to prove that a modal
logic characterizes $\pbis$ over reactive probabilistic processes, it is enough to show that it can
discriminate all reactive probabilistic trees of \emph{finite} height. A specific condition on the depth of
distinguishing formulas has also to be satisfied, where $\depth(\phi)$ is defined as usual:
\[
\begin{array}{rcl}
\depth(\top) & \!\!\! = \!\!\! & 0 \\
\depth(\lnot \phi') & \!\!\! = \!\!\! & \depth(\phi') \\
\depth(\phi_{1} \land \phi_{2}) & \!\!\! = \!\!\! & \max(\depth(\phi_{1}), \depth(\phi_{2})) \\
\depth(\phi_{1} \lor \phi_{2}) & \!\!\! = \!\!\! & \max(\depth(\phi_{1}), \depth(\phi_{2})) \\
\depth(\diam{a}{p} \phi') & \!\!\! = \!\!\! & 1 + \depth(\phi') \\
\end{array}
\]

	\begin{proposition}\label{prop:log_char_scaling}

Let $\call$ be one of the probabilistic modal logics in Sect.~\ref{sec:pml}. If $\call$ characterizes $=$
over $\RPTf$ and for any two nodes $t_{1}$ and $t_{2}$ of an arbitrary $\RPTf$ model such that $t_{1} \neq
t_{2}$ there exists $\phi \in \call$ distinguishing $t_{1}$ from $t_{2}$ such that:
\[
\depth(\phi) \: \le \: \max(\height(t_{1}), \height(t_{2}))
\]
then $\call$ characterizes $\pbis$ over the set of RPLTS models.

		\begin{proof}
Given two states $s_{1}$ and $s_{2}$ of an RPLTS, if $s_{1} \pbis s_{2}$ then for all $n \in \natns$ it
holds that $\prune{n}{\dsb{s_{1}}} = \prune{n}{\dsb{s_{2}}}$ thanks to Cor.~\ref{cor:indprinc}, hence
$s_{1}$ and $s_{2}$ satisfy the same formulas of $\call$ because $\call$ characterizes $=$ over $\RPTf$.
Suppose now that $s_{1} \not\pbis s_{2}$ and consider the minimum $n \in \natns_{\ge 1}$ for which
$\prune{n}{\dsb{s_{1}}} \neq \prune{n}{\dsb{s_{2}}}$. Then there exists $\phi \in \call$ distinguishing
$\prune{n}{\dsb{s_{1}}}$ from $\prune{n}{\dsb{s_{2}}}$ such that $\depth(\phi) \le
\max(\height(\prune{n}{\dsb{s_{1}}}), \height(\prune{n}{\dsb{s_{2}}})) = n$, hence the same formula $\phi$
also distinguishes $s_{1}$ from $s_{2}$.
\fullbox

		\end{proof}

	\end{proposition}

Notice that, in the proof above, if $\depth(\phi)$ were greater than $n$ then, in general, $\phi$ may not
distinguish higher prunings of $\dsb{s_{1}}$ and $\dsb{s_{2}}$, nor may any formula of depth at most $n$ and
derivable from~$\phi$ still distinguish $\prune{n}{\dsb{s_{1}}}$ from $\prune{n}{\dsb{s_{2}}}$.

	\begin{example}\label{ex:depth_ngt_height}

Consider a process whose initial state $s_{1}$ has only an $a$-transition to a state having only a
$c$-transition to $\nil$, and another process whose initial state $s_{2}$ has only a $b$-transition to a
state having only a $d$-transition to $\nil$. Their corresponding trees differ at height $n = 1$ because
$\prune{1}{\dsb{s_{1}}}$ has an $a$-transition to $\nil$ while $\prune{1}{\dsb{s_{2}}}$ has a $b$-transition
to $\nil$.

The formula of depth $2$ given by $\diam{a}{1} \lnot \diam{c}{1}$ distinguishes $\prune{1}{\dsb{s_{1}}}$
from $\prune{1}{\dsb{s_{2}}}$, but this is no longer the case with $\prune{2}{\dsb{s_{1}}}$ and
$\prune{2}{\dsb{s_{2}}}$ as neither satisfies that formula.

The formula of depth $2$ given by $\diam{a}{1} \lor \diam{b}{1} \diam{c}{1}$ distinguishes
$\prune{1}{\dsb{s_{1}}}$ from $\prune{1}{\dsb{s_{2}}}$, but this is no longer the case with the derived
formula $\diam{a}{1} \lor \diam{b}{1}$ of depth $1$ as both nodes satisfy it.
\fullbox

	\end{example}

Based on the considerations above, in this section we show the main result of the paper: the logical
equivalence induced by \PMLo\ has the same discriminating power as $\pbis$.

This result is accomplished in three steps. Firstly, we redemonstrate Larsen and Skou's result for \PMLna\
in the $\RPTf$ setting, which yields a proof that, with respect to the one in~\cite{LS91}, is simpler and
does not require the minimal deviation assumption (i.e., that the probability associated with any state in
the support of the target distribution of a transition be a multiple of some value). This provides a proof
scheme for the subsequent steps. Secondly, we demonstrate that \PMLno\ characterizes $\pbis$ by adapting the
proof scheme to cope with the replacement of conjunction with disjunction. Thirdly, we demonstrate that
\PMLo\ characterizes $\pbis$ by further adapting the proof scheme to cope with the absence of negation. 

Moreover, we redemonstrate Desharnais, Edalat, and Panangaden's result for \PMLa\ through yet another
adaptation of the proof scheme that, unlike the proof in~\cite{DEP02}, is constructive and works directly on
\emph{discrete} state spaces without making use of measure-theoretic arguments based on analytic spaces.
Avoiding the resort to measure theory was shown to be possible for the first time by Worrell in an
unpublished note cited in~\cite{Pan11}.

\subsection{\PMLna\ Characterizes $\pbis$: A New Proof}
\label{sec:pmlna_pbis}

To show that the logical equivalence induced by \PMLna\ implies node equality $=$, we reason on the
contrapositive. Given two nodes $t_{1}$ and $t_{2}$ such that $t_{1} \neq t_{2}$, we proceed by induction on
the height of $t_{1}$ to find a distinguishing \PMLna\ formula whose depth is not greater than the heights
of $t_{1}$ and $t_{2}$. The idea is to exploit negation, so to ensure that certain distinguishing formulas
are \emph{satisfied} by a certain derivative $t'$ of $t_{1}$ (rather than the derivatives of $t_{2}$
different from $t'$), then take the \emph{conjunction} of those formulas preceded by a diamond decorated
with the probability for $t_{1}$ of \emph{reaching} $t'$.

The only non-trivial case is the one in which $t_{1}$ and $t_{2}$ enable the same actions. At least one of
those actions, say $a$, is such that, after performing it, the two nodes reach two distributions $\Delta_{1,
a}$ and $\Delta_{2, a}$ such that $\Delta_{1, a} \neq \Delta_{2, a}$. Given a node $t' \in \supp(\Delta_{1,
a})$ such that $\Delta_{1, a}(t') > \Delta_{2, a}(t')$, by the induction hypothesis there exists a \PMLna\
formula $\phi'_{2, j}$ that distinguishes $t'$ from a specific $t'_{2, j} \in \supp(\Delta_{2, a}) \setminus
\{ t' \}$. We can assume that $t' \models \phi'_{2, j} \not\rightmodels t'_{2, j}$ otherwise, thanks to the
presence of negation in \PMLna, it would suffice to consider $\lnot\phi'_{2, j}$.

As a consequence, $t_{1} \models \diam{a}{\Delta_{1, a}(t')} \bigwedge_{j} \phi'_{2, j} \not\rightmodels
t_{2}$ because $\Delta_{1, a}(t') > \Delta_{2, a}(t')$ and $\Delta_{2, a}(t')$ is the maximum probabilistic
lower bound for which $t_{2}$ satisfies a formula of that form. Notice that $\Delta_{1, a}(t')$ may not be
the maximum probabilistic lower bound for which $t_{1}$ satisfies such a formula, because $\bigwedge_{j}
\phi'_{2, j}$ might be satisfied by other $a$-derivatives of $t_{1}$ in $\supp(\Delta_{1, a}) \setminus \{
t' \}$.

	\begin{theorem}\label{thm:pmlna_pbis}

Let $(T, A, \! \trs{} \!)$ be in $\RPTf$ and $t_{1}, t_{2} \in T$. Then $t_{1} = t_{2}$ iff $t_{1} \models
\phi \iff t_{2} \models \phi$ for all $\phi \in \PMLna$. Moreover, if $t_{1} \neq t_{2}$, then there exists
$\phi \in \PMLna$ distinguishing $t_{1}$ from $t_{2}$ such that $\depth(\phi) \le \max(\height(t_{1}),
\height(t_{2}))$.

		\begin{proof}
Given $t_{1}, t_{2} \in T$, we proceed as follows:

			\begin{itemize}

\item If $t_{1} = t_{2}$, then obviously $t_{1} \models \phi \iff t_{2} \models \phi$ for all $\phi \in
\PMLna$.

\item Assuming that $t_{1} \neq t_{2}$, we show that there exists $\phi \in \PMLna$, with $\depth(\phi) \le
\max(\height(t_{1}), \height(t_{2}))$, such that it is not the case that $t_{1} \models \phi \iff t_{2}
\models \phi$ by proceeding by induction on $\height(t_{1}) \in \natns$:

				\begin{itemize}

\item If $\height(t_{1}) = 0$, then $\height(t_{2}) \ge 1$ because $t_{1} \neq t_{2}$. As a consequence,
$t_{2}$ has at least one outgoing transition, say labeled with $a$, hence $t_{1} \not\models \diam{a}{1}
\rightmodels t_{2}$. Notice that $\depth(\diam{a}{1}) = 1 \le \max(\height(t_{1}), \height(t_{2}))$.

\item Let $\height(t_{1}) = n + 1$ for some $n \in \natns$ and suppose that for all $t'_{1}, t'_{2} \in T$
such that $t'_{1} \neq t'_{2}$ and $\height(t'_{1}) \le n$ there exists $\phi' \in \PMLna$, with
$\depth(\phi') \le \max(\height(t'_{1}), \height(t'_{2}))$, such that it is not the case that $t'_{1}
\models \phi' \iff t'_{2} \models \phi'$. Let $\init(t_{h})$, $h \in \{ 1, 2 \}$, be the set of actions in
$A$ labeling the transitions departing from $t_{h}$:

					\begin{itemize}

\item If $\init(t_{1}) \neq \init(t_{2})$, then it holds that $t_{1} \models \diam{a}{1} \not\rightmodels
t_{2}$ for some $a \in \init(t_{1}) \setminus \init(t_{2})$ or $t_{1} \not\models \diam{a}{1} \rightmodels
t_{2}$ for some $a \in \init(t_{2}) \setminus \init(t_{1})$. Notice that $\depth(\diam{a}{1}) = 1 \le
\max(\height(t_{1}), \height(t_{2}))$.

\item If $\init(t_{1}) = \init(t_{2})$, then $\init(t_{1}) \neq \emptyset \neq \init(t_{2})$ as
$\height(t_{1}) \ge 1$. Since $t_{1} \neq t_{2}$, there must exist $a \in \init(t_{1})$ such that $t_{1}
\trs{a} \Delta_{1, a}$, $t_{2} \trs{a} \Delta_{2, a}$, and $\Delta_{1, a} \neq \Delta_{2, a}$. From
$\Delta_{1, a} \neq \Delta_{2, a}$, it follows that there exists $t' \in \supp(\Delta_{1, a})$ such that $1
\ge \Delta_{1, a}(t') > \Delta_{2, a}(t') \ge 0$. Assuming that $\supp(\Delta_{2, a}) \backslash \{ t' \} =
\{ t'_{2, 1}, t'_{2, 2}, \dots, t'_{2, k} \}$, which cannot be empty because there must also exist $t'_{2}
\in \supp(\Delta_{2, a})$ such that $0 \le \Delta_{1, a}(t'_{2}) < \Delta_{2, a}(t'_{2}) \le 1$, by the
induction hypothesis for each $j = 1, 2, \dots, k$ there exists $\phi'_{2, j} \in \PMLna$, with
$\depth(\phi'_{2, j}) \le \max(\height(t'), \linebreak \height(t'_{2, j}))$, such that it is not the case
that $t' \models \phi'_{2, j} \iff t'_{2, j} \models \phi'_{2, j}$. Since \PMLna\ includes negation, without
loss of generality we can assume that $t' \models \phi'_{2, j} \not\rightmodels t'_{2, j}$. Therefore, it
holds that $t_{1} \models \diam{a}{\Delta_{1, a}(t')} \bigwedge_{1 \le j \le k} \phi'_{2, j}
\not\rightmodels t_{2}$ because $\Delta_{1, a}(t') > \Delta_{2, a}(t')$ and $\Delta_{2, a}(t')$ is the
maximum probabilistic lower bound for which $t_{2}$ satisfies a formula of that form. Notice that the
resulting formula, which we denote by $\phi$ for short, satisfies:
\cws{8}{\begin{array}{l}
\hspace*{-0.9cm} \depth(\phi) = 1 + \max_{1 \le j \le k} \depth(\phi'_{2, j}) \\
\hspace*{-0.5cm} \le 1 + \max_{1 \le j \le k} \max(\height(t'), \height(t'_{2, j})) \\
\hspace*{-0.5cm} = 1 + \max(\height(t'), \max_{1 \le j \le k} \height(t'_{2, j})) \\
\hspace*{-0.5cm} = \max(1 + \height(t'), 1 + \max_{1 \le j \le k} \height(t'_{2,j})) \\
\hspace*{-0.5cm} \le \max(\height(t_{1}), \height(t_{2})) \\
\end{array}}
\fullbox

					\end{itemize}

				\end{itemize}

			\end{itemize}

		\end{proof}

	\end{theorem}

\vspace{-1ex}

\subsection{\PMLno\ Characterizes $\pbis$: Adapting the Proof}
\label{sec:pmlno_pbis}
\vspace{-0.5ex}

Since $\phi_{1} \land \phi_{2}$ is logically equivalent to $\lnot (\lnot \phi_{1} \lor \lnot \phi_{2})$, it
is not surprising that \PMLno\ characterizes $\pbis$ too. However, the proof of this result will be useful
to set up an outline of the proof of the main result of this paper, i.e., that \PMLo\ characterizes $\pbis$
as well.

Similarly to Thm.~\ref{thm:pmlna_pbis}, also for \PMLno\ we reason on the contrapositive and
proceed by induction. Given $t_{1},t_{2}$ such that $t_{1} \neq t_{2}$, we intend to exploit negation,
so to ensure that certain distinguishing formulas are \emph{not satisfied} by a certain derivative $t'$ of
$t_{1}$ (rather than the derivatives of~$t_{2}$ different from $t'$), then take the \emph{disjunction} of
those formulas preceded by a diamond decorated with the probability for $t_{2}$ of \emph{not reaching} $t'$.
The only non-trivial case is $t' \in \supp(\Delta_{1, a})$ such that $\Delta_{1, a}(t') > \Delta_{2,
a}(t')$.  By inductive hypothesis there exists a formula $\phi'_{2, j}$ distinguishing $t'$
from a specific $t'_{2, j} \in \supp(\Delta_{2, a}) \setminus \{ t' \}$. We can assume that $t' \not\models
\phi'_{2, j} \rightmodels t'_{2, j}$ (otherwise we consider $\lnot\phi'_{2, j}$ since negation is in \PMLno). 
Therefore, $t_{1} \not\models \diam{a}{1 - \Delta_{2, a}(t')}
\bigvee_{j} \phi'_{2, j} \rightmodels t_{2}$ because $1 - \Delta_{2, a}(t') > 1 - \Delta_{1, a}(t')$ and the
maximum probabilistic lower bound for which $t_{1}$ satisfies a formula of that form cannot exceed $1 -
\Delta_{1, a}(t')$. Notice that $1 - \Delta_{2, a}(t')$ is the \emph{maximum} lower bound for
which $t_{2}$ satisfies such a formula, because that value is the probability with which $t_{2}$ does not
reach $t'$ after performing $a$.

	\begin{theorem}\label{thm:pmlno_pbis}

Let $(T, A, \! \trs{} \!)$ be in $\RPTf$ and $t_{1}, t_{2} \in T$. Then $t_{1} = t_{2}$ iff $t_{1} \models
\phi \iff t_{2} \models \phi$ for all $\phi \in \PMLno$. Moreover, if $t_{1} \neq t_{2}$, then there exists
$\phi \in \PMLno$ distinguishing $t_{1}$ from $t_{2}$ such that $\depth(\phi) \le \max(\height(t_{1}),
\height(t_{2}))$.

		\begin{proof}
The proof is similar to the one of Thm.~\ref{thm:pmlna_pbis}, apart from the final part of the last subcase,
which changes as follows. \\
By the induction hypothesis, for each $j = 1, 2, \dots, k$ there exists $\phi'_{2, j} \! \in \PMLno$, with
$\depth(\phi'_{2, j}) \le \max(\height(t'), \height(t'_{2, j}))$, such that it is not the case that $t'
\models \phi'_{2, j} \iff t'_{2, j} \models \phi'_{2, j}$. Since \PMLno\ includes negation, without loss of
generality we can assume that $t' \not\models \phi'_{2, j} \rightmodels t'_{2, j}$. Therefore, it holds that
$t_{1} \not\models \diam{a}{1 - \Delta_{2, a}(t')} \bigvee_{1 \le j \le k} \phi'_{2, j} \rightmodels t_{2}$
because $1 - \Delta_{2, a}(t') > 1 - \Delta_{1, a}(t')$ and the maximum probabilistic lower bound for which
$t_{1}$ satisfies a formula of that form cannot exceed $1 - \Delta_{1, a}(t')$.
\fullbox

		\end{proof}

	\end{theorem}

\subsection{Also \PMLo\ Characterizes $\pbis$}
\label{sec:pmlo_pbis}

The proof that \PMLo\ characterizes $\pbis$ is inspired by the one for \PMLno, thus considers the
contrapositive and proceeds by induction. In the only non-trivial case, we will arrive at a situation in
which $t_{1} \not\models \diam{a}{1 - (\Delta_{2, a}(t') + p)} \bigvee_{j \in J} \phi'_{2, j} \rightmodels
t_{2}$ for:

	\begin{itemize}

\item a derivative $t'$ of $t_{1}$, such that $\Delta_{1, a}(t') > \Delta_{2, a}(t')$, not satisfying any
subformula $\phi'_{2, j}$;

\item a suitable probabilistic value $p$ such that $\Delta_{2, a}(t') + p < 1$;

\item an index set $J$ identifying certain derivatives of~$t_{2}$ other than $t'$.

	\end{itemize}

The choice of $t'$ is crucial, because negation is no longer available in \PMLo. Unlike the case of \PMLno,
this induces the introduction of~$p$ and the limitation to~$J$ in the format of the distinguishing formula.
An important observation is that, in many cases, a disjunctive distinguishing formula can be obtained from a
conjunctive one by suitably \emph{increasing} some probabilistic lower bounds.

	\begin{figure*}[t]

\centerline{\includegraphics{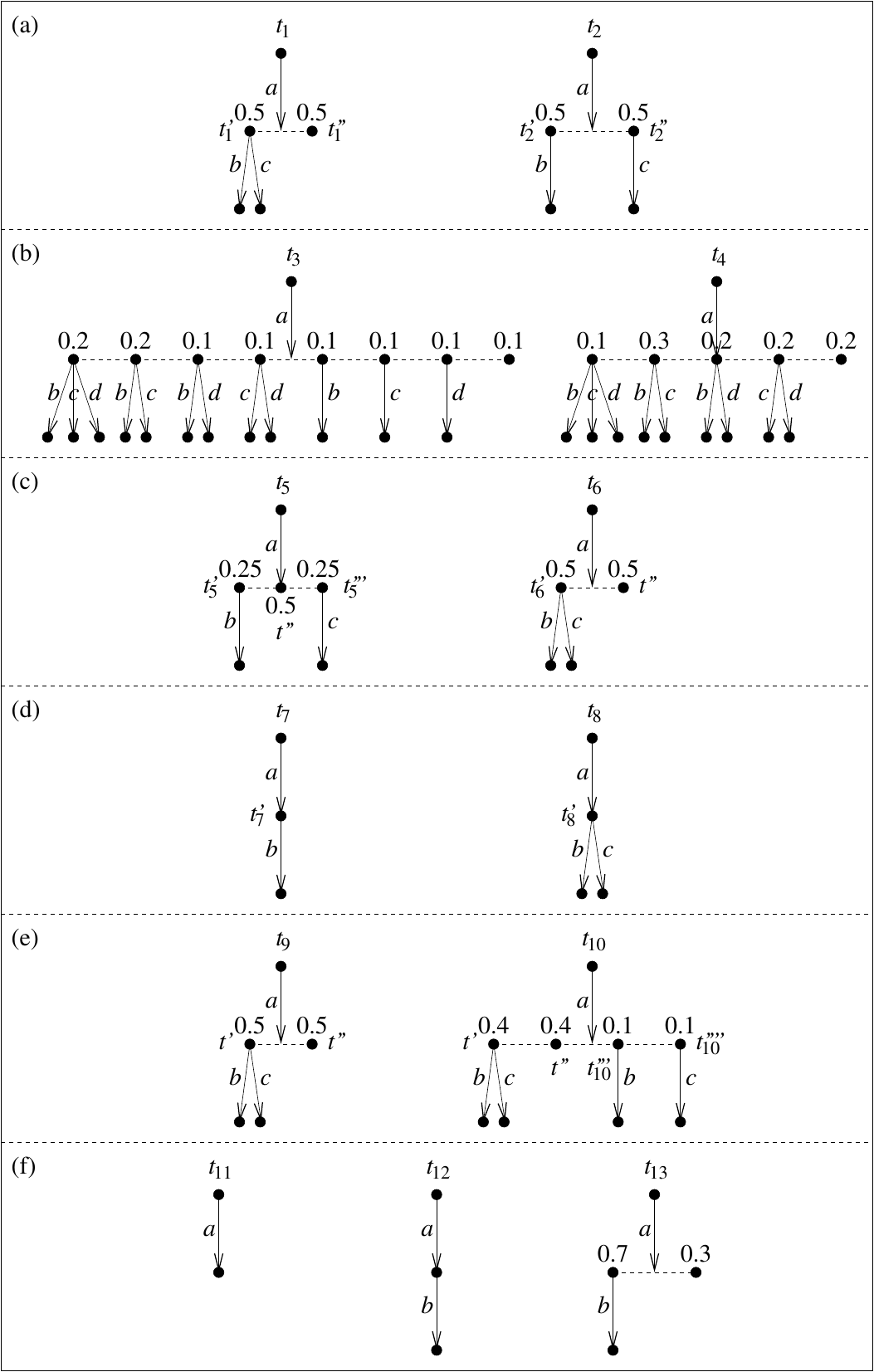}}
\caption{$\RPTf$ models used in the examples of Sects.~\ref{sec:pmlo_pbis} and~\ref{sec:pmla_pbis}}
\label{fig:pml_or_examples}

	\end{figure*}

	\begin{example}\label{ex:increase_from_and_to_or}

The nodes $t_{1}$ and $t_{2}$ in Fig.~\ref{fig:pml_or_examples}(a) cannot be distinguished by any formula in
which neither conjunction nor \linebreak disjunction occurs. It holds that:
\[
\begin{array}{rcccl}
t_{1} & \!\!\! \models \!\!\! & \diam{a}{0.5} \, (\diam{b}{1} \land \diam{c}{1}) & \!\!\! \not\rightmodels
\!\!\! & t_{2} \\
t_{1} & \!\!\! \not\models \!\!\! & \diam{a}{1.0} \, (\diam{b}{1} \lor \diam{c}{1}) & \!\!\! \rightmodels
\!\!\! & t_{2} \\
\end{array}
\]
Notice that, when moving from the conjunctive formula to the disjunctive one, the probabilistic lower bound
decorating the \linebreak $a$-diamond increases from $0.5$ to $1$ and the roles of $t_{1}$ and $t_{2}$ with
respect to $\models$ are inverted.

The situation is similar for the nodes $t_{3}$ and $t_{4}$ in Fig.~\ref{fig:pml_or_examples}(b), where two
occurrences of conjunction/disjunction are necessary:
\[
\begin{array}{rcccl}
t_{3} & \!\!\! \models \!\!\! & \diam{a}{0.2} \, (\diam{b}{1} \land \diam{c}{1} \land \diam{d}{1}) & \!\!\!
\not\rightmodels \!\!\! & t_{4} \\
t_{3} & \!\!\! \models \!\!\! & \diam{a}{0.9} \, (\diam{b}{1} \lor \diam{c}{1} \lor \diam{d}{1}) & \!\!\!
\not\rightmodels \!\!\! & t_{4} \\
\end{array}
\]
but the roles of $t_{3}$ and $t_{4}$ with respect to $\models$ cannot be inverted.
\fullbox

	\end{example}

However, increasing some of the probabilistic lower bounds in a conjunctive distinguishing formula does not
always yield a disjunctive one. This is the case when the use of conjunction/disjunction is not necessary
for telling two different nodes apart.

	\begin{example}\label{ex:no_increase}

For the nodes $t_{5}$ and $t_{6}$ in Fig.~\ref{fig:pml_or_examples}(c), it holds that:
\[
t_{5} \: \not\models \: \diam{a}{0.5} \, (\diam{b}{1} \land \diam{c}{1}) \: \rightmodels \: t_{6}
\]
If we replace conjunction with disjunction and we vary the probabilistic lower bound between $0.5$ and $1$,
we produce no disjunctive formula capable of discriminating between $t_{5}$ and $t_{6}$. Nevertheless, a
distinguishing formula belonging to \PMLo\ exists because:
\[
t_{5} \: \not\models \: \diam{a}{0.5} \, \diam{b}{1} \: \rightmodels \: t_{6}
\]
where disjunction does not occur at all.
\fullbox

	\end{example}

The examples above show that the increase of some probabilistic lower bounds when moving from conjunctive
distinguishing formulas to disjunctive ones takes place only in the case that the probabilities of reaching
certain nodes have to be \emph{summed up}. Additionally, we recall that, in order for two nodes to be
related by $\pbis$, they must enable the same actions, so focussing on a \emph{single} action is enough for
discriminating when only disjunction is available.

Bearing this in mind, for any node $t$ of finite height we define the set $\phior(t)$ of \PMLo\ formulas
satisfied by $t$ featuring:

	\begin{itemize}

\item probabilistic lower bounds of diamonds that are \emph{maximal} with respect to the satisfiability of a
formula of that format by $t$ \linebreak (this is consistent with the observation in the last sentence
\linebreak before Thm.~\ref{thm:pmlno_pbis}, and keeps the set $\phior(t)$ finite);

\item diamonds that arise only from \emph{existing} transitions that depart from $t$ (so to avoid useless
diamonds in disjunctions and hence keep the set $\phior(t)$ finite);

\item disjunctions that stem only from \emph{single} transitions of \emph{different} nodes in the support of
a distribution reached by $t$ (transitions departing from the same node would result in formulas like
$\bigvee_{h \in H} \diam{a_{h}}{p_{h}} \phi_{h}$, with $a_{h_{1}} \neq a_{h_{2}}$ for $h_{1} \neq h_{2}$,
which are useless for discriminating with respect to $\pbis$) and are preceded by a diamond decorated with
the \emph{sum} of the probabilities assigned to those nodes by the distribution reached by $t$.

	\end{itemize}

	\begin{definition}\label{def:phi_or_set}

The set $\phior(t)$ for a node $t$ of finite height is defined by induction on $\height(t)$ as follows:

		\begin{itemize}

\item If $\height(t) = 0$, then $\phior(t) = \emptyset$.

\item If $\height(t) \ge 1$ for $t$ having transitions of the form $t \trs{a_{i}} \Delta_{i}$ with
$\supp(\Delta_{i}) = \{ t'_{i, j} \mid j \in J_{i} \}$ and $i \in I \neq \emptyset$, then:
\[
\begin{array}{rcl}
\phior(t) & \!\!\! = \!\!\! & \{ \diam{a_{i}}{1} \mid i \in I \} \\
& \!\!\! \cup \!\!\! & \bigcup\limits_{i \in I} \hplb(\bigcup\limits_{\emptyset \neq J' \subseteq J_{i}} \{
\diam{a_{i}}{\sum\limits_{j \in J'} \Delta_{i}(t'_{i, j})} \bigvee\limits_{j \in J'}^{.} \phi'_{i, j, k}
\mid \\[0.4cm]
& & \hspace*{1.6cm} t'_{i, j} \in \supp(\Delta_{i}), \phi'_{i, j, k} \in \phior(t'_{i, j}) \}) \\
\end{array}
\]
where operator $\dot{\lor}$ is a variant of $\lor$ in which identical operands are not admitted (i.e.,
idempotence is forced) and function $\hplb$ keeps only the formula with the highest probabilistic lower
bound decorating the initial $a_{i}$-diamond among the formulas differring only for that bound.
\fullbox

		\end{itemize}

	\end{definition}

To illustrate the definition given above, we exhibit some examples showing the usefulness of $\phior$-sets
for discrimination purposes. In particular, let us reconsider the non-trivial case mentioned at the
beginning of this subsection. Given two different nodes that with the same action reach two different
distributions, a good criterion for choosing $t'$ (a derivative of the first node not satisfying certain
formulas, to which the first distribution assigns a probability greater than the second one) seems to be the
\emph{minimality} of the $\phior$-set.

	\begin{example}\label{ex:no_nondet_choice}

For the nodes $t_{7}$ and $t_{8}$ in Fig.~\ref{fig:pml_or_examples}(d), we have:
\[
\begin{array}{rcl}
\phior(t_{7}) & \!\!\! = \!\!\! & \{ \diam{a}{1}, \, \diam{a}{1} \diam{b}{1} \} \\
\phior(t_{8}) & \!\!\! = \!\!\! & \{ \diam{a}{1}, \, \diam{a}{1} \diam{b}{1}, \, \diam{a}{1} \diam{c}{1} \}
\\
\end{array}
\]
A formula like $\diam{a}{1} \, (\diam{b}{1} \lor \diam{c}{1})$ is useless for discriminating between $t_{7}$
and $t_{8}$, because disjunction is between two actions enabled by the same node and hence constituting a
nondeterministic choice. Indeed, such a formula is not part of $\phior(t_{8})$. While in the case of
conjunction it is often necessary to concentrate on several alternative actions, in the case of disjunction
it is convenient to focus on a single action per node when aiming at producing a distinguishing formula.

The fact that $\diam{a}{1} \diam{c}{1} \in \phior(t_{8})$ is a distinguishing formula can be retrieved as
follows. Starting from the two identically labeled transitions $t_{7} \trs{a} \Delta_{7, a}$ and $t_{8}
\trs{a} \Delta_{8, a}$ where $\Delta_{7, a}(t'_{7}) = 1 = \Delta_{8, a}(t'_{8})$ and $\Delta_{7, a}(t'_{8})
= 0 = \Delta_{8, a}(t'_{7})$, we have:
\[
\begin{array}{rcl}
\phior(t'_{7}) & \!\!\! = \!\!\! & \{ \diam{b}{1} \} \\
\phior(t'_{8}) & \!\!\! = \!\!\! & \{ \diam{b}{1}, \, \diam{c}{1} \} \\
\end{array}
\]
If we focus on $t'_{7}$ because $\Delta_{7, a}(t'_{7}) > \Delta_{8, a}(t'_{7})$ and its $\phior$-set is
minimal, then $t'_{7} \not\models \diam{c}{1} \rightmodels t'_{8}$ with $\diam{c}{1} \in \phior(t'_{8})
\setminus \phior(t'_{7})$. As a consequence, $t_{7} \not\models \diam{a}{1} \diam{c}{1} \rightmodels t_{8}$
where the value $1$ decorating the $a$-diamond stems from $1 - \Delta_{8, a}(t'_{7})$.
\fullbox

	\end{example}

	\begin{example}\label{ex:same_basic_formulas_sum_prob}

For the nodes $t_{1}$ and $t_{2}$ in Fig.~\ref{fig:pml_or_examples}(a), we have:
\[
\begin{array}{rcl}
\phior(t_{1}) & \!\!\! = \!\!\! & \{ \diam{a}{1}, \, \diam{a}{0.5} \diam{b}{1}, \, \diam{a}{0.5} \diam{c}{1}
\} \\
\phior(t_{2}) & \!\!\! = \!\!\! & \{ \diam{a}{1}, \, \diam{a}{0.5} \diam{b}{1}, \, \diam{a}{0.5}
\diam{c}{1}, \, \diam{a}{1} \, (\diam{b}{1} \lor \diam{c}{1}) \} \\
\end{array}
\]
The formulas with two diamonds and no disjunction are identical in the two sets, so their disjunction
$\diam{a}{0.5} \diam{b}{1} \lor \diam{a}{0.5} \diam{c}{1}$ is useless for discriminating between $t_{1}$ and
$t_{2}$. Indeed, such a formula is part of neither $\phior(t_{1})$ nor $\phior(t_{2})$. In contrast, their
disjunction in which decorations of identical diamonds are summed up, i.e., $\diam{a}{1} \, (\diam{b}{1}
\lor \diam{c}{1})$, is fundamental. It belongs only to $\phior(t_{2})$ because in the case of $t_{1}$ the
$b$-transition and the $c$-transition depart from the same node, hence no probabilities can be added.

The fact that $\diam{a}{1} \, (\diam{b}{1} \lor \diam{c}{1}) \in \phior(t_{2})$ is a distinguishing formula
can be retrieved as follows. Starting from the two identically labeled transitions $t_{1} \trs{a} \Delta_{1,
a}$ and $t_{2} \trs{a} \Delta_{2, a}$ where $\Delta_{1, a}(t'_{1}) = \Delta_{1, a}(t''_{1}) = 0.5 =
\Delta_{2, a}(t'_{2}) = \Delta_{2, a}(t''_{2})$ and $\Delta_{1, a}(t'_{2}) = \Delta_{1, a}(t''_{2}) = 0 =
\Delta_{2, a}(t'_{1}) = \Delta_{2, a}(t''_{1})$, we have:
\[
\begin{array}{rcl@{\qquad}rcl}
\phior(t'_{1}) & \!\!\! = \!\!\! & \{ \diam{b}{1}, \, \diam{c}{1} \} &
\phior(t''_{1}) & \!\!\! = \!\!\! & \emptyset \\
\phior(t'_{2}) & \!\!\! = \!\!\! & \{ \diam{b}{1} \} &
\phior(t''_{2}) & \!\!\! = \!\!\! & \{ \diam{c}{1} \} \\
\end{array}
\]
If we focus on $t''_{1}$ because $\Delta_{1, a}(t''_{1}) > \Delta_{2, a}(t''_{1})$ and its $\phior$-set is
minimal, then $t''_{1} \not\models \diam{b}{1} \rightmodels t'_{2}$ with $\diam{b}{1} \in \phior(t'_{2})
\setminus \phior(t''_{1})$ as well as $t''_{1} \not\models \diam{c}{1} \rightmodels t''_{2}$ with
$\diam{c}{1} \in \phior(t''_{2}) \setminus \phior(t''_{1})$. As a consequence, $t_{1} \not\models
\diam{a}{1} \, (\diam{b}{1} \lor \diam{c}{1}) \rightmodels t_{2}$ where the value $1$ decorating the
$a$-diamond stems from $1 - \Delta_{2, a}(t''_{1})$.
\fullbox

	\end{example}

	\begin{example}\label{ex:diff_basic_formulas_same_prob}

For the nodes $t_{5}$ and $t_{6}$ in Fig.~\ref{fig:pml_or_examples}(c), we have:
\[
\hspace*{-0.1cm}\begin{array}{rcl}
\phior(t_{5}) & \!\!\!\! = \!\!\!\! & \{ \diam{a}{1}, \, \diam{a}{0.25} \diam{b}{1}, \, \diam{a}{0.25}
\diam{c}{1}, \, \diam{a}{0.5} \, (\diam{b}{1} \lor \diam{c}{1}) \} \\
\phior(t_{6}) & \!\!\!\! = \!\!\!\! & \{ \diam{a}{1}, \, \diam{a}{0.5} \diam{b}{1}, \, \diam{a}{0.5}
\diam{c}{1} \} \\
\end{array}
\]
The formulas with two diamonds and no disjunction are different in the two sets, so they are enough for
discriminating between $t_{5}$ and~$t_{6}$. In contrast, the only formula with disjunction, which belongs to
$\phior(t_{5})$, is useless because the probability decorating its $a$-diamond is equal to the probability
decorating the $a$-diamond of each of the two formulas with two diamonds in $\phior(t_{6})$.

The fact that $\diam{a}{0.5} \diam{b}{1} \in \phior(t_{6})$ is a distinguishing formula can be retrieved as
follows. Starting from the two identically labeled transitions $t_{5} \trs{a} \Delta_{5, a}$ and $t_{6}
\trs{a} \Delta_{6, a}$ where $\Delta_{5, a}(t'_{5}) = \Delta_{5, a}(t'''_{5}) = 0.25$, $\Delta_{5, a}(t'') =
0.5 = \Delta_{6, a}(t'_{6}) = \Delta_{6, a}(t'')$, and $\Delta_{5, a}(t'_{6}) = 0 = \Delta_{6, a}(t'_{5}) =
\Delta_{6, a}(t'''_{5})$, we have:
\[
\begin{array}{rcl@{\qquad}rcl}
\phior(t'_{5}) & \!\!\! = \!\!\! & \{ \diam{b}{1} \} &
\phior(t'''_{5}) & \!\!\! = \!\!\! & \{ \diam{c}{1} \} \\
\phior(t'_{6}) & \!\!\! = \!\!\! & \{ \diam{b}{1}, \, \diam{c}{1} \} &
\phior(t'') & \!\!\! = \!\!\! & \emptyset \\
\end{array}
\]
Notice that $t''$ might be useless for discriminating purposes because it has the same probability in both
distributions, so we exclude it. If we focus on $t'''_{5}$ because $\Delta_{5, a}(t'''_{5}) > \Delta_{6,
a}(t'''_{5})$ and its $\phior$-set is minimal after the exclusion of $t''$, then $t'''_{5} \not\models
\diam{b}{1} \rightmodels t'_{6}$ with $\diam{b}{1} \in \phior(t'_{6}) \setminus \phior(t'''_{5})$, while no
distinguishing formula is considered with respect to $t''$ as element of $\supp(\Delta_{6, a})$ due to the
exclusion of $t''$ itself. As a consequence, $t_{5} \not\models \diam{a}{0.5} \diam{b}{1} \rightmodels
t_{6}$ where the value $0.5$ decorating the $a$-diamond stems from $1 - (\Delta_{6, a}(t'''_{5}) + p)$ with
$p = \Delta_{6, a}(t'')$. The reason for subtracting the probability that $t_{6}$ reaches $t''$ after
performing $a$ is that $t'' \not\models \diam{b}{1}$.

We conclude by observing that focussing on $t''$ as derivative with the minimum $\phior$-set is indeed
problematic, because it would result in $\diam{a}{0.5} \diam{b}{1}$ when considering $t''$ as derivative of
$t_{5}$, but it would result in $\diam{a}{0.5} \, (\diam{b}{1} \lor \diam{c}{1})$ when considering $t''$ as
derivative of $t_{6}$, with the latter formula not distinguishing between $t_{5}$ and $t_{6}$. Moreover,
when focussing on $t'''_{5}$, no formula $\phi'$ could have been found such that $t'''_{5} \not\models \phi'
\rightmodels t''$ as $\phior(t'') \subsetneq \phior(t'''_{5})$.
\fullbox

	\end{example}

The last example shows that, in the general format for the \PMLo\ distinguishing formula mentioned at the
beginning of this subsection, i.e., $\diam{a}{1 - (\Delta_{2, a}(t') + p)} \bigvee_{j \in J} \phi'_{2, j}$,
the set $J$ only contains any derivative of the second node different from $t'$ to which the two
distributions assign two \emph{different} probabilities. No derivative of the two original nodes having the
same probability in both distributions is taken into account even if its $\phior$-set is minimal -- because
it might be useless for discriminating purposes -- nor is it included in~$J$ -- because there might be no
formula satisfied by this node when viewed as a derivative of the second node, which is not satisfied by
$t'$. Furthermore, the value $p$ is the probability that the second node reaches the excluded derivatives
that do \emph{not} satisfy $\bigvee_{j \in J} \phi'_{2, j}$; note that the first node reaches those
derivatives with the same probability $p$.

We present two additional examples illustrating some technicalities of Def.~\ref{def:phi_or_set}. The former
example shows the usefulness of the operator~$\dot{\lor}$ and of the function $\hplb$ for selecting the
right $t'$ on the basis of the minimality of its $\phior$-set among the derivatives of the first node to
which the first distribution assigns a probability greater than the second one. The latter example
emphasizes the role played, for the same purpose as before, by formulas occurring in a $\phior$-set whose
number of nested diamonds is not maximal.

	\begin{example}\label{ex:dot_or_hplb}

For the nodes $t_{9}$ and $t_{10}$ in Fig.~\ref{fig:pml_or_examples}(e), we have:
\[
\hspace*{-0.1cm}\begin{array}{rcl}
\phior(t_{9}) & \!\!\! = \!\!\! & \{ \diam{a}{1}, \, \diam{a}{0.5} \diam{b}{1}, \, \diam{a}{0.5} \diam{c}{1}
\} \\
\phior(t_{10}) & \!\!\! = \!\!\! & \{ \diam{a}{1}, \, \diam{a}{0.5} \diam{b}{1}, \, \diam{a}{0.5}
\diam{c}{1}, \, \diam{a}{0.6} \, (\diam{b}{1} \lor \diam{c}{1}) \} \\
\end{array}
\]
Starting from the two identically labeled transitions $t_{9} \trs{a} \Delta_{9, a}$ and $t_{10} \trs{a}
\Delta_{10, a}$ where $\Delta_{9, a}(t') = \Delta_{9, a}(t'') = 0.5$, $\Delta_{10, a}(t') \linebreak =
\Delta_{10, a}(t'') = 0.4$, $\Delta_{10, a}(t'''_{10}) = \Delta_{10, a}(t''''_{10}) = 0.1$, and $\Delta_{9,
a}(t'''_{10}) = \Delta_{9, a}(t''''_{10}) = 0$, we have:
\[
\begin{array}{rcl@{\qquad}rcl}
\phior(t') & \!\!\! = \!\!\! & \{ \diam{b}{1}, \, \diam{c}{1} \} &
\phior(t'') & \!\!\! = \!\!\! & \emptyset \\
\phior(t'''_{10}) & \!\!\! = \!\!\! & \{ \diam{b}{1} \} &
\phior(t''''_{10}) & \!\!\! = \!\!\! & \{ \diam{c}{1} \} 
\end{array}
\]
If we focus on $t''$ because $\Delta_{9, a}(t'') > \Delta_{10, a}(t'')$ and its $\phior$-set is minimal,
then $t'' \not\models \diam{b}{1} \rightmodels t'$ with $\diam{b}{1} \in \phior(t') \setminus \phior(t'')$,
$t'' \not\models \diam{b}{1} \rightmodels t'''_{10}$ with $\diam{b}{1} \in \phior(t'''_{10}) \setminus
\phior(t'')$, and $t'' \not\models \diam{c}{1} \rightmodels t''''_{10}$ with $\diam{c}{1} \in
\phior(t''''_{10}) \setminus \phior(t'')$. As a consequence, $t_{9} \not\models \diam{a}{0.6} \,
(\diam{b}{1} \lor \diam{c}{1}) \rightmodels t_{10}$ where the formula belongs to $\phior(t_{10})$ and the
value $0.6$ decorating the $a$-diamond stems from $1 - \Delta_{10, a}(t'')$.

If $\lor$ were used in place of $\dot{\lor}$, then in $\phior(t_{10})$ we would also have formulas like
$\diam{a}{0.5} \, (\diam{b}{1} \lor \diam{b}{1})$ and $\diam{a}{0.5} \, (\diam{c}{1} \lor \diam{c}{1})$.
These are useless in that logically equivalent to other formulas already in $\phior(t_{10})$ in which
disjunction does not occur and, most importantly, would apparently augment the size of $\phior(t_{10})$, an
inappropriate fact in the case that $t_{10}$ were a derivative of some other node instead of being the root
of a tree.

If $\hplb$ were not used, then in $\phior(t_{10})$ we would also have formulas like $\diam{a}{0.1}
\diam{b}{1}$, $\diam{a}{0.4} \diam{b}{1}$, $\diam{a}{0.1} \diam{c}{1}$, and $\diam{a}{0.4} \diam{c}{1}$, in
which the probabilistic lower bounds of the $a$-diamonds are not maximal with respect to the satisfiability
of formulas of that form by $t_{10}$; those with maximal probabilistic lower bounds associated with
$a$-diamonds are $\diam{a}{0.5} \diam{b}{1}$ and $\diam{a}{0.5} \diam{c}{1}$, which already belong to
$\phior(t_{10})$. In the case that $t_{9}$ and $t_{10}$ were derivatives of two nodes under comparison
instead of being the roots of two trees, the presence of those additional formulas in $\phior(t_{10})$ may
lead to focus on $t_{10}$ instead of $t_{9}$ -- for reasons that will be clear in
Ex.~\ref{ex:formulas_without_or} -- thereby producing no distinguishing formula.
\fullbox

	\end{example}

	\begin{example}\label{ex:no_max_diam_nesting}

For the nodes $t_{11}$, $t_{12}$, $t_{13}$ in Fig.~\ref{fig:pml_or_examples}(f), we have:
\[
\begin{array}{rcl}
\phior(t_{11}) & \!\!\! = \!\!\! & \{ \diam{a}{1} \} \\
\phior(t_{12}) & \!\!\! = \!\!\! & \{ \diam{a}{1}, \, \diam{a}{1} \diam{b}{1} \} \\
\phior(t_{13}) & \!\!\! = \!\!\! & \{ \diam{a}{1}, \, \diam{a}{0.7} \diam{b}{1} \} \\
\end{array}
\]
Let us view them as derivatives of other nodes, rather than roots of trees. The presence of formula
$\diam{a}{1}$ in $\phior(t_{12})$ and $\phior(t_{13})$ -- although it has not the maximum number of nested
diamonds in those two sets -- ensures the minimality of $\phior(t_{11})$ and hence that $t_{11}$ is selected
for building a distinguishing formula. If $\diam{a}{1}$ were not in $\phior(t_{12})$ and $\phior(t_{13})$,
then $t_{12}$ and $t_{13}$ could be selected, but no distinguishing formula satisfied by $t_{11}$ could be
obtained.
\fullbox

	\end{example}

The criterion for selecting the right $t'$ based on the minimality of its $\phior$-set has to take into
account a further aspect related to \emph{formulas without disjunctions}. If two derivatives -- with
different probabilities in the two distributions -- have the same formulas without disjunctions in their
$\phior$-sets, then a distinguishing formula for the two nodes will have disjunctions in it (see
Exs.~\ref{ex:same_basic_formulas_sum_prob} and~\ref{ex:dot_or_hplb}). In contrast, if the formulas without
disjunctions are different between the two $\phior$-sets, then one of those formulas will tell the two
derivatives apart (see Ex.~\ref{ex:no_nondet_choice}).

A particular instance of the second case is the one in which for each formula without disjunctions in one of
the two $\phior$-sets there is a variant in the other $\phior$-set -- i.e., a formula without disjunctions
that has the same format but may differ for the values of some probabilistic lower bounds -- and vice versa.
In this event, \emph{regardless of the minimality} of the $\phior$-sets, it has to be selected the
derivative such that (i)~for each formula without disjunctions in its $\phior$-set there exists a variant in
the $\phior$-set of the other derivative such that the probabilistic lower bounds in the former formula are
$\le$ than the corresponding bounds in the latter formula and (ii)~at least one probabilistic lower bound in
a formula without disjunctions in the $\phior$-set of the selected derivative is $<$ than the corresponding
bound in the corresponding variant in the $\phior$-set of the other derivative. We say that the $\phior$-set
of the selected derivative is a \emph{$(\le, <)$-variant} of the $\phior$-set of the other derivative.

	\begin{example}\label{ex:formulas_without_or}

Let us view the nodes $t_{5}$ and $t_{6}$ in Fig.~\ref{fig:pml_or_examples}(c) as derivatives of other
nodes, rather than roots of trees. Based on their $\phior$-sets shown in
Ex.~\ref{ex:diff_basic_formulas_same_prob}, we should focus on $t_{6}$ because $\phior(t_{6})$ contains
fewer formulas. However, by so doing, we would be unable to find a distinguishing formula in $\phior(t_{5})$
that is not satisfied by $t_{6}$. Indeed, if we look carefully at the formulas without disjunctions in
$\phior(t_{5})$ and $\phior(t_{6})$, we note that they differ only for their probabilistic lower bounds:
$\diam{a}{1} \in \phior(t_{6})$ is a variant of $\diam{a}{1} \in \phior(t_{5})$, $\diam{a}{0.5} \diam{b}{1}
\in \phior(t_{6})$ is a variant of $\diam{a}{0.25} \diam{b}{1} \in \phior(t_{5})$, and $\diam{a}{0.5}
\diam{c}{1} \in \phior(t_{6})$ is a variant of $\diam{a}{0.25} \diam{c}{1} \in \phior(t_{5})$. \linebreak
Therefore, we must focus on $t_{5}$ because $\phior(t_{5})$ contains formulas without disjunctions such as
$\diam{a}{0.25} \diam{b}{1}$ and $\diam{a}{0.25} \diam{c}{1}$ having smaller bounds: $\phior(t_{5})$ is a
$(\le, <)$-variant of $\phior(t_{6})$.

Consider now the nodes $t_{9}$ and $t_{10}$ in Fig.~\ref{fig:pml_or_examples}(e), whose $\phior$-sets are
shown in Ex.~\ref{ex:dot_or_hplb}. If function $\hplb$ were not used and hence $\phior(t_{10})$ also
contained $\diam{a}{0.1} \diam{b}{1}$, $\diam{a}{0.4} \diam{b}{1}$, $\diam{a}{0.1} \diam{c}{1}$, and
$\diam{a}{0.4} \diam{c}{1}$, then the formulas without disjunctions in $\phior(t_{9})$ would no longer be
equal to those in $\phior(t_{10})$. More precisely, the formulas without disjunctions would be similar
between the two sets, with those in $\phior(t_{10})$ having smaller probabilistic lower bounds, so that we
would erroneously focus on $t_{10}$.
\fullbox

	\end{example}

Summing up, in the construction of the \PMLo\ distinguishing formula mentioned at the beginning of this
subsection, i.e., $\diam{a}{1 - (\Delta_{2, a}(t') + p)} \bigvee_{j \in J} \phi'_{2, j}$, the steps for
choosing the derivative~$t'$, on the basis of which each subformula $\phi'_{2, j}$ is then generated so that
it is not satisfied by $t'$, are the following:

	\begin{enumerate}

\item Consider only derivatives to which $\Delta_{1, a}$ assigns a probability greater than the one assigned
by $\Delta_{2, a}$.

\item Within the previous set, eliminate all the derivatives whose \linebreak $\phior$-sets have $(\le,
<)$-variants.

\item Among the remaining derivatives, focus on one of those having a minimal $\phior$-set.

	\end{enumerate}

	\begin{theorem}\label{thm:pmlo_pbis}

Let $(T, A, \! \trs{} \!)$ be in $\RPTf$ and $t_{1}, t_{2} \in T$. Then $t_{1} = t_{2}$ iff $t_{1} \models
\phi \iff t_{2} \models \phi$ for all $\phi \in \PMLo$. Moreover, if $t_{1} \neq t_{2}$, then there exists
$\phi \in \PMLo$ distinguishing $t_{1}$ from $t_{2}$ such that $\depth(\phi) \le \max(\height(t_{1}),
\height(t_{2}))$.

		\begin{proof}
Given $t_{1}, t_{2} \in T$, we proceed as follows:

			\begin{itemize}

\item If $t_{1} = t_{2}$, then obviously $t_{1} \models \phi \iff t_{2} \models \phi$ for all $\phi \in
\PMLo$.

\item Assuming that $t_{1} \neq t_{2}$, we show that there exists $\phi \in \phior(t_{1}) \cup
\phior(t_{2})$, which ensures that $\depth(\phi) \le \max(\height(t_{1}), \linebreak \height(t_{2}))$, such
that it is not the case that $t_{1} \models \phi \iff t_{2} \models \phi$ by proceeding by induction on
$\height(t_{1}) \in \natns$. The proof is similar to the one of Thm.~\ref{thm:pmlno_pbis}, in particular in
the cases $\height(t_{1}) = 0$ and $\height(t_{1}) = n + 1$ with $\init(t_{1}) \neq \init(t_{2})$ it
benefits from the presence of $\{ \diam{a_{i}}{1} \mid i \in I \}$ in $\phior(t)$ as of
Def.~\ref{def:phi_or_set}. However, it changes as follows before the application of the induction hypothesis
in the case $\height(t_{1}) = n + 1$ with $\init(t_{1}) = \init(t_{2}) \neq \emptyset$ and $t_{1} \trs{a}
\Delta_{1, a}$, $t_{2} \trs{a} \Delta_{2, a}$, and $\Delta_{1, a} \neq \Delta_{2, a}$ for some $a \in
\init(t_{1})$. \\
Let $\supp_{a} = \supp(\Delta_{1, a}) \cup \supp(\Delta_{2, a})$, which can be partitioned into $\supp_{a,
\neq} = \{ t' \in \supp_{a} \mid \Delta_{1, a}(t') \neq \Delta_{2, a}(t') \}$ and $\supp_{a, =} = \{ t' \in
\supp_{a} \mid \Delta_{1, a}(t') = \Delta_{2, a}(t') \}$ with $|\supp_{a, \neq}| \ge 2$ because $\Delta_{1,
a} \neq \Delta_{2, a}$ and $|\supp_{a, =}| \ge 0$. \linebreak We recall that $\phior(t'')$ is a $(\le,
<)$-variant of $\phior(t')$ iff:

				\begin{itemize}

\item for each formula without disjunctions in one of the two $\phior$-sets there is a variant in the other
$\phior$-set -- i.e., a formula without disjunctions that has the same format but may differ for the values
of some probabilistic lower bounds -- and vice versa (this means that there exists a bijection between the
formulas without disjunctions in the two $\phior$-sets, because the maximality of the probabilistic lower
bounds in a $\phior$-set implies the existence of at most one formula with a given format in the
$\phior$-set);

\item for each formula without disjunctions in $\phior(t'')$ there exists a variant in $\phior(t')$ such
that the probabilistic lower bounds in the former formula are $\le$ than the corresponding bounds in the
latter formula;

\item at least one probabilistic lower bound in a formula without disjunctions in $\phior(t'')$ is $<$ than
the corresponding bound in the corresponding variant in $\phior(t')$.

				\end{itemize}

Among all the nodes in $\supp_{a, \neq}$, there exists one denoted by $t'$ such that, for all $t'' \in
\supp_{a, \neq} \setminus \{ t' \}$, $\phior(t'')$ is not a $(\le, <)$-variant of $\phior(t')$ as we now
prove by proceeding by induction on $|\supp_{a, \neq}| \in \natns_{\ge 2}$:

				\begin{itemize}

\item If $|\supp_{a, \neq}| = 2$ -- hence $\supp_{a, \neq} = \{ t', t'' \}$ -- then trivially at least one
of $\phior(t')$ and $\phior(t'')$ is not a $(\le, <)$-variant of the other.

\item Let $|\supp_{a, \neq}| = n + 1$ for some $n \in \natns_{\ge 2}$ and suppose that the result holds for
each subset of $\supp_{a, \neq}$ of cardinality between $2$ and $n$. Assuming that $\supp_{a, \neq} = \{
t'_{1}, t'_{2}, \dots, t'_{n + 1} \}$, we denote by $t'$ the node in $\supp_{a, \neq} \setminus
\{ t'_{n + 1} \}$ that, by the induction hypothesis, enjoys the property over that subset. There are two
cases:

					\begin{itemize}

\item If $\phior(t'_{n + 1})$ is not a $(\le, <)$-variant of $\phior(t')$ either, then $t'$ enjoys the
property over the entire set $\supp_{a, \neq}$.

\item Suppose that $\phior(t'_{n + 1})$ is a $(\le, <)$-variant of $\phior(t')$, which implies that
$\phior(t')$ cannot be a $(\le, <)$-variant of $\phior(t'_{n + 1})$. From the fact that, for all $t'' \in
\supp_{a, \neq} \setminus \{ t', \linebreak t'_{n + 1} \}$, $\phior(t'')$ is not a $(\le, <)$-variant of
$\phior(t')$ by the induction hypothesis, it follows that $\phior(t'')$ is not a $(\le, <)$-variant of
$\phior(t'_{n + 1})$. Indeed, for each such $t''$ the set $\phior(t'')$ contains at least a formula without
disjunctions that is not a variant of any formula without disjunctions in $\phior(t')$, or all formulas
without disjunctions in $\phior(t'')$ are identical to formulas without disjunctions in $\phior(t')$, hence
this holds true with respect to $\phior(t'_{n + 1})$ too, given that $\phior(t'_{n + 1})$ is a $(\le,
<)$-variant of $\phior(t')$. As a consequence, $t'_{n + 1}$ enjoys the property over the entire set
$\supp_{a, \neq}$.

					\end{itemize}

				\end{itemize}

Within the set of all the nodes in $\supp_{a, \neq}$ enjoying the property above, we select one with a
minimal $\phior$-set, which we denote by $t'_{\rm min}$. Suppose that $\Delta_{1, a}(t'_{\rm min}) >
\Delta_{2, a}(t'_{\rm min})$ and let $t'_{2, j}$ be an arbitrary node belonging to $\supp_{a, \neq, 2} =
(\supp_{a, \neq} \setminus \{ t'_{\rm min} \}) \cap \supp(\Delta_{2, a})$. By the induction hypothesis, from
$t'_{2, j} \neq t'$ it follows that there exists $\phi'_{2, j} \in \phior(t'_{\rm min}) \cup \phior(t'_{2,
j})$ such that it is not the case that $t'_{\rm min} \models \phi'_{2, j} \iff t'_{2, j} \models \phi'_{2,
j}$. In particular, it holds that $t'_{\rm min} \not\models \phi'_{2, j} \rightmodels t'_{2, j}$ because
$\phi'_{2, j} \in \phior(t'_{2, j})$, as can be seen by considering the following two cases based on the
fact that $\phior(t'_{2, j})$ is not a $(\le, <)$-variant of $\phior(t'_{\rm min})$:

				\begin{itemize}

\item If at least one formula without disjunctions in $\phior(t'_{2, j})$ is not a variant of any formula
without disjunctions in $\phior(t'_{\rm min})$, then such a formula can be taken as $\phi'_{2, j}$ given the
maximality of the probabilistic lower bounds of any basic formula in $\phior(t'_{\rm min})$.

\item If all basic formulas in $\phior(t'_{2, j})$ are identical to basic formulas in $\phior(t'_{\rm
min})$, then $\phior(t'_{2, j})$ must contain some more formulas (with disjunctions) not in $\phior(t'_{\rm
min})$ given the minimality of the latter set, otherwise we would have selected $t'_{2, j}$ in place of
$t'_{\rm min}$. One of the additional formulas (with disjunctions) in $\phior(t'_{2, j})$ can be taken as
$\phi'_{2, j}$.

				\end{itemize}

Letting $\supp_{a, =, \not\models} = \{ t' \in \supp_{a, =} \mid t' \not\models \bigvee_{t'_{2, j} \in
\supp_{a, \neq, 2}} \phi'_{2, j} \}$ as well as $p_{\not\models} = \Delta_{2, a}(\supp_{a, =, \not\models})
= \Delta_{1, a}(\supp_{a, =, \not\models})$, we have that $t_{1} \not\models \diam{a}{1 - (\Delta_{2,
a}(t'_{\rm min}) + p_{\not\models})} \bigvee_{t'_{2, j} \in \supp_{a, \neq, 2}} \phi'_{2, j} \rightmodels
t_{2}$ because $1 - (\Delta_{2, a}(t'_{\rm min}) + p_{\not\models}) > 1 - (\Delta_{1, a}(t'_{\rm min}) +
p_{\not\models})$ and the maximum probabilistic lower bound for which $t_{1}$ satisfies a formula of that
form cannot exceed $1 - (\Delta_{1, a}(t'_{\rm min}) + p_{\not\models})$. \linebreak The \PMLo\
distinguishing formula above may not be in $\phior(t_{2})$, but it is logically implied by, or equivalent
to, a distinguishing formula in $\phior(t_{2})$ for the following reasons:

				\begin{itemize}

\item Each $t'_{2, j}$ belongs to $\supp(\Delta_{2, a})$.

\item Each $\phi'_{2, j}$ belongs to $\phior(t'_{2, j})$.

\item The probabilistic lower bound $1 - (\Delta_{2, a}(t'_{\rm min}) + p_{\not\models})$ is equal to
$\sum_{t'_{2, j} \in \supp_{a, \neq, 2}} \Delta_{2, a}(t'_{2, j}) + \Delta_{2, a}(\supp_{a, =, \models})$,
so in the \PMLo\ distinguishing formula it is sufficient to replace $\bigvee_{t'_{2, j} \in \supp_{a, \neq,
2}} \phi'_{2, j}$ with $\dot{\bigvee}_{t' \in \supp_{a, \neq, 2} \cup \supp_{a, =, \models}} \phi_{t'}$
where:

					\begin{itemize}

\item $\phi_{t'} = \phi'_{2, j}$ if $t' = t'_{2, j}$ for some $j$;

\item $\phi_{t'} = \phi' \in \phior(t')$ if $t' \neq t'_{2, j}$ for all $j$, where $\phi' \implies \phi'_{2,
j}$ for some $j$ and the existence of such a $\phi'$ in $\phior(t')$ stems from $t' \in \supp_{a, =,
\models}$, i.e., $t' \models \bigvee_{t'_{2, j} \in \supp_{a, \neq, 2}} \phi'_{2, j}$.
\fullbox

					\end{itemize}

				\end{itemize}

			\end{itemize}

		\end{proof}

	\end{theorem}

\subsection{\PMLa\ Characterizes $\pbis$: A Direct Proof for Discrete Systems}
\label{sec:pmla_pbis}

By adapting the proof of Thm.~\ref{thm:pmlo_pbis} consistently with the proof of Thm.~\ref{thm:pmlna_pbis},
in our setting we can also demonstrate that \PMLa\ characterizes $\pbis$ by working directly on
\emph{discrete} state spaces. The idea is to obtain $t_{1} \models \diam{a}{\Delta_{1, a}(t') + p}
\bigwedge_{j \in J} \phi'_{2, j} \not\rightmodels t_{2}$.

To this purpose, for any node $t$ of finite height we define the set $\phiand(t)$ of \PMLa\ formulas
satisfied by $t$ featuring, in addition to maximal probabilistic lower bounds and diamonds arising only from
transitions of $t$ as for $\phior(t)$, conjunctions that (i) stem only from transitions departing from the
\emph{same node} in the support of a distribution reached by $t$ and (ii) are preceded by a diamond
decorated with the \emph{sum} of the probabilities assigned by that distribution to that node and other
nodes with the \emph{same transitions} considered for that node. Formally, given $t$ having transitions of
the form $t \trs{a_{i}} \Delta_{i}$ with $\supp(\Delta_{i}) = \{ t'_{i, j} \mid j \in J_{i} \}$ and $i \in I
\neq \emptyset$, we let:
\[
\begin{array}{rcl}
\phiand(t) & \!\!\! = \!\!\! & \{ \diam{a_{i}}{1} \mid i \in I \} \\[0.1cm]
& \!\!\! \cup \!\!\! & \bigcup\limits_{i \in I} \splb(\lmp \diam{a_{i}}{\Delta_{i}(t'_{i, j})}
\bigwedge\limits_{k \in K'} \phi'_{i, j, k} \mid \emptyset \! \neq \! K' \! \subseteq \! K_{i, j}, \\[0.3cm]
& & \hspace*{1.7cm} t'_{i, j} \in \supp(\Delta_{i}), \phi'_{i, j, k} \in \phiand(t'_{i, j}) \rmp) \\
\end{array}
\]
where $\lmp$ and $\rmp$ are multiset parentheses, $K_{i, j}$ is the index set for $\phiand(t'_{i, j})$, and
function $\splb$ merges all formulas possibly differring only for the probabilistic lower bound decorating
their initial $a_{i}$-diamond by summing up those bounds (notice that such formulas stem from different
nodes in $\supp(\Delta_{i})$).

We now provide some examples illustrating the technicalities of the definition above, as well as the fact
that a good criterion for choosing $t'$ occurring in the \PMLa\ distinguishing formula at the beginning of
this subsection is the \emph{maximality} of the $\phiand$-set.

	\begin{example}\label{ex:multiset_splb}

In Fig.~\ref{fig:pml_or_examples}(b), the multiset giving rise to $\phiand(t_{3})$ contains two occurrences
of $\diam{a}{0.2} \diam{b}{1}$ and two occurrences of $\diam{a}{0.1} \diam{b}{1}$, which are merged into
$\diam{a}{0.6} \diam{b}{1}$ by function $\splb$. Likewise, the multiset behind $\phiand(t_{4})$ contains
formulas $\diam{a}{0.1} \diam{b}{1}$, $\diam{a}{0.3} \diam{b}{1}$, and $\diam{a}{0.2} \diam{b}{1}$, which
are merged into $\diam{a}{0.6} \diam{b}{1}$.
\fullbox

	\end{example}

	\begin{example}\label{ex:phi_and_maximality}

For the nodes $t_{1}$ and $t_{2}$ in Fig.~\ref{fig:pml_or_examples}(a), we have:
\[
\begin{array}{rcl}
\phiand(t_{1}) & \!\!\! = \!\!\! & \{ \diam{a}{1}, \, \diam{a}{0.5} \diam{b}{1}, \, \diam{a}{0.5}
\diam{c}{1}, \, \diam{a}{0.5} \, (\diam{b}{1} \land \diam{c}{1}) \} \\
\phiand(t_{2}) & \!\!\! = \!\!\! & \{ \diam{a}{1}, \, \diam{a}{0.5} \diam{b}{1}, \, \diam{a}{0.5}
\diam{c}{1} \} \\
\end{array}
\]
The conjunction $\diam{a}{0.5} \diam{b}{1} \land \diam{a}{0.5} \diam{c}{1}$ is useless for discriminating
between $t_{1}$ and $t_{2}$ -- it is part of neither $\phiand(t_{1})$ nor $\phiand(t_{2})$ -- while
$\diam{a}{0.5} \, (\diam{b}{1} \land \diam{c}{1})$ is the only distinguishing formula and belongs only to
$\phiand(t_{1})$, because in the case of $t_{2}$ the $b$-transition and the $c$-transition depart from two
different nodes. Starting from the two identically labeled transitions $t_{1} \trs{a} \Delta_{1,
a}$ and $t_{2} \trs{a} \Delta_{2, a}$ where $\Delta_{1, a}(t'_{1}) = \Delta_{1, a}(t''_{1}) = 0.5 =
\Delta_{2, a}(t'_{2}) = \Delta_{2, a}(t''_{2})$ and $\Delta_{1, a}(t'_{2}) = \Delta_{1, a}(t''_{2}) = 0 =
\Delta_{2, a}(t'_{1}) = \Delta_{2, a}(t''_{1})$, we have:
\[
\begin{array}{rcl@{\qquad}rcl}
\phiand(t'_{1}) & \!\!\! = \!\!\! & \{ \diam{b}{1}, \, \diam{c}{1} \} &
\phiand(t''_{1}) & \!\!\! = \!\!\! & \emptyset \\
\phiand(t'_{2}) & \!\!\! = \!\!\! & \{ \diam{b}{1} \} &
\phiand(t''_{2}) & \!\!\! = \!\!\! & \{ \diam{c}{1} \} \\
\end{array}
\]
If we focus on $t'_{1}$ because $\Delta_{1, a}(t'_{1}) > \Delta_{2, a}(t'_{1})$ and its $\phiand$-set is
maximal, then $t'_{1} \models \diam{c}{1} \not\rightmodels t'_{2}$ with $\diam{c}{1} \in \phiand(t'_{1})
\setminus \phiand(t'_{2})$ as well as $t'_{1} \models \diam{b}{1} \not\rightmodels t''_{2}$ with
$\diam{b}{1} \in \phiand(t'_{1}) \setminus \phiand(t''_{2})$. As a consequence, $t_{1} \models \diam{a}{0.5}
\, (\diam{b}{1} \land \diam{c}{1}) \not\rightmodels t_{2}$ where the value $0.5$ decorating the $a$-diamond
stems from $\Delta_{1, a}(t'_{1})$.
\fullbox

	\end{example}

As far as the other two variables occurring in the \PMLa\ distinguishing formula at the beginning of this
subsection are concerned, $J$ only contains any derivative of the second node different from $t'$ to which
the two distributions assign two \emph{different} probabilities, while $p$ is the probability of reaching
derivatives having the \emph{same} probability in both distributions that \emph{satisfy} $\bigwedge_{j \in
J} \phi'_{2, j}$. Moreover, when selecting $t'$, we have to leave out all the derivatives whose
$\phiand$-sets have $(\le, <)$-variants.

	\begin{theorem}\label{thm:pmla_pbis}

Let $(T, A, \! \trs{} \!)$ be in $\RPTf$ and $t_{1}, t_{2} \in T$. Then $t_{1} = t_{2}$ iff $t_{1} \models
\phi \iff t_{2} \models \phi$ for all $\phi \in \PMLa$. Moreover, if $t_{1} \neq t_{2}$, then there exists
$\phi \in \PMLa$ distinguishing $t_{1}$ from $t_{2}$ such that $\depth(\phi) \le \max(\height(t_{1}),
\height(t_{2}))$.

		\begin{proof}
Similar to that of Thm.~\ref{thm:pmlo_pbis}, with these differences:

			\begin{itemize}

\item We select $t'_{\rm max}$ as one of the nodes with maximal $\phiand$-set in~$\supp_{a, \neq}$ having no
$(\le, <)$-variants.

\item It holds that $t'_{\rm max} \models \phi'_{2, j} \not\rightmodels t'_{2, j}$ for all $t'_{2, j} \in
\supp_{a, \neq, 2}$ because $\phi'_{2, j} \in \phiand(t'_{\rm max})$ thanks to the maximality of
$\phiand(t'_{\rm max})$.

\item Letting $\supp_{a, =, \models} = \{ t' \in \supp_{a, =} \mid t' \models \bigwedge_{t'_{2, j} \in
\supp_{a, \neq, 2}} \phi'_{2, j} \}$ as well as $p_{\models} = \Delta_{1, a}(\supp_{a, =, \models}) =
\Delta_{2, a}(\supp_{a, =, \models})$, we have that $t_{1} \models \diam{a}{\Delta_{1, a}(t'_{\rm max}) +
p_{\models}} \bigwedge_{t'_{2, j} \in \supp_{a, \neq, 2}} \phi'_{2, j} \not\rightmodels t_{2}$ \linebreak
because $\Delta_{1, a}(t'_{\rm max}) + p_{\models} > \Delta_{2, a}(t'_{\rm max}) + p_{\models}$ and the
maximum probabilistic lower bound for which $t_{2}$ satisfies a formula of that form cannot exceed
$\Delta_{2, a}(t'_{\rm max}) + p_{\models}$.

\item The \PMLa\ distinguishing formula is in $\phiand(t_{1})$ due to $\splb$.
\fullbox

			\end{itemize}

		\end{proof}

	\end{theorem}

%% file: concl.tex
\section{Conclusions}
\label{sec:concl}

In this paper, we have studied modal logic characterizations of bisimilarity over reactive probabilistic
processes. Starting from previous work by Larsen and Skou~\cite{LS91} (who provided a characterization based
on an extension of Hennessy-Milner modal logic where diamonds are decorated with probabilistic lower bounds)
and by Desharnais, Edalat, and Panangaden~\cite{DEP02} (who showed that negation is not necessary), we have
proved that conjunction can be replaced by disjunction without having to reintroduce negation. Thus, in the
reactive probabilistic setting, conjunction and disjunction are \emph{interchangeable} to characterize
(bi)simulation equivalence, while they are \emph{both necessary} for simulation preorder~\cite{DGJP03}. As a
side result, using the same proof technique we have provided alternative proofs of the expressiveness of the
logics \PMLna\ and \PMLa.

The intuition behind our result for \PMLo\ is that from a conjunctive distinguishing formula it is often possible to derive a disjunctive one by suitably increasing some probabilistic lower bounds. 
This corresponds to summing up the probabilities of reaching certain states that are in the support of a target distribution.
In fact, a state of an RPLTS can be given a semantics as a reactive probabilistic tree, and hence it is characterized by the countable set of formulas (approximated by the $\phior$-set) obtained by doing finite visits of the tree.

On the application side, the \PMLo-based characterization of bisimilarity over reactive probabilistic processes may help to prove a conjecture in~\cite{BSV14b} about the discriminating power of three different testing equivalences respectively using reactive probabilistic tests, fully nondeterministic tests, and nondeterministic and probabilistic tests. 
Many examples lead to conjecture that testing equivalence based on nondeterministic and probabilistic tests has the same discriminating power as bisimilarity.
Given two $\pbis$-inequivalent reactive probabilistic processes, the idea of the tentative proof is to build a distinguishing nondeterministic and probabilistic test from a distinguishing \PMLa\ formula.
One of the main difficulties with carrying out such a proof, i.e., the fact that choices within tests fit well together with disjunction rather than conjunction, may be overcome by starting from a distinguishing \PMLo\ formula.

\vspace{-1ex}